\documentclass[english,prl,aps,twocolumn,10pt]{revtex4-2}

\usepackage{amsmath}
\usepackage{amssymb}
\usepackage{graphicx}
\usepackage{babel}
\usepackage{array}
\usepackage{verbatim}
\usepackage[colorlinks=true, pdfstartview=FitV, linkcolor=blue, citecolor=blue, urlcolor=blue]{hyperref} % enable links

\def\ie       {{\it i.e.}}

\newcommand{\ee}[1]{\cdot10^{#1}}
\newcommand{\mr}[1]{\mathrm{#1}}
\newcommand{\unit}[1]{\,\mathrm{#1}}
\newcommand{\um}{\,\mu{\rm m}}

\newcommand{\ye}{\gamma_\mr{e}}

\newcommand{\Bip}{B_\mr{ip}}

\newcommand{\Bx}{B_{\bar x}}
\newcommand{\By}{B_{\bar y}}
\newcommand{\Bz}{B_{\bar z}}
\newcommand{\Bmin}{B_\mr{min}}

\newcommand{\vecB}{\vec B}
\newcommand{\Isat}{I_\mr{sat}}
\newcommand{\SNR}{\mr{SNR}}
\newcommand{\Sx}{\hat{S}_{\bar x}}
\newcommand{\Sy}{\hat{S}_{\bar y}}
\newcommand{\Sz}{\hat{S}_{\bar z}}

\newcommand{\eps}{\epsilon}
\newcommand{\Dw}{\Delta \omega}

\newcommand{\wpm}{\omega_{\pm}}
\newcommand{\wa}{\omega_{+}}
\newcommand{\wb}{\omega_{-}}

\begin{document}

\title{Scanning nitrogen-vacancy center magnetometry in large in-plane magnetic fields}

\author{P.~Welter$^1$, J.~Rhensius$^2$, A.~Morales$^2$, M.~S.~Wörnle$^{1,3}$, C.-H.~Lambert$^3$, G.~Puebla-Hellmann$^2$, P.~Gambardella$^{3,4}$ and C.~L.~Degen$^{1,4}$}
\email{degenc@ethz.ch}
\affiliation{$^1$Department of Physics, ETH Zurich, Otto Stern Weg 1, 8093 Zurich, Switzerland.}
\affiliation{$^2$QZabre AG, Regina-Kagi-Strasse 11, 8050 Zurich, Switzerland}
\affiliation{$^3$Department of Material Science, ETH Zurich, Honggerbergring 64, 8093 Zurich, Switzerland.}
\affiliation{$^4$Quantum Center, ETH Zurich, 8093 Zurich, Switzerland.}

\begin{abstract}
Scanning magnetometry with nitrogen-vacancy (NV) centers in diamond has emerged as a powerful microscopy for studying weak stray field patterns with nanometer resolution.  Due to the internal crystal anisotropy of the spin defect, however, external bias fields -- critical for the study of magnetic materials -- must be applied along specific spatial directions.   In particular, the most common diamond probes made from $\{100\}$-cut diamond only support fields at an angle of $\theta = 55^\circ$ from the surface normal.  In this paper, we report fabrication of scanning diamond probes from $\{110\}$-cut diamond where the spin anisotropy axis lies in the scan plane ($\theta=90^\circ$).  We show that these probes retain their sensitivity in large in-plane fields and demonstrate scanning magnetometry of the domain pattern of Co-NiO films in applied fields up to 40\,mT. Our work extends scanning NV magnetometry to the important class of materials that require large in-plane fields.
\end{abstract}

\date{\today}

\maketitle

%\textit{Introduction -- }
%
The imaging of magnetic fields with nanometer resolution is of key importance to existing and emerging nanotechnologies and offers a unique view on many physical phenomena and processes.  For example, the analysis of small ferromagnetic or antiferromagnetic features, such as domains walls or junctions, is crucial for the development of next-generation data storage technology and spintronic devices \cite{kosub17,appel19,wornle19}. The imaging of stray fields also provides access to novel materials and phases, including skyrmions \cite{dovzhenko18}, ferroelectrics \cite{gross17}, complex oxides \cite{ariyaratne18}, superconductors \cite{thiel16}, and topological insulators \cite{nowack13}.  Nanoscale magnetic imaging can further be used to detect current flow in nanoscale conductors, with applications to semiconductor physics, and integrated circuits \cite{chang17, ku20}.  Provided sufficient detection sensitivity and spatial resolution, the measurement of nanoscale magnetic fields can therefore provide a detailed, and often complementary, view on the structure and function of nanoscale systems.

Scanning magnetometers based on single spins in diamond have recently emerged as a powerful method for the quantitative imaging of nanoscale stray fields.  Exploiting the principles of quantum metrology, scanning NV magnetometers combine excellent sensitivity with  high spatial resolution, opening the door for studying weak magnetic features that are difficult to access with existing magnetic probes.  The technique has, for example, enabled imaging of ferromagnetic vortices and domain walls~\cite{rondin13,tetienne14,tetienne15}, helimagnetic~\cite{dussaux16} and skyrmionic systems~\cite{dovzhenko18,gross18,jenkins19}, two-dimensional ferromagnets~\cite{thiel19,sun21,fabre21}, as well as ferrimagnetic~\cite{velez19}, antiferromagnetic~\cite{appel19,wornle19,wornle21,hedrich21} and multiferroic materials~\cite{gross17}.  The technique has also been applied to investigate magnetic excitations and spin waves~\cite{wolfe14,vandersar15,bertelli20}.
Despite of the rapid progress, scanning NV magnetometry has been mostly confined to zero or weak bias fields ($\lesssim 5\unit{mT}$), or in some cases, to larger fields applied along a tilted angle with respect to the sample surface.  This limitation is mainly due to the internal crystal anisotropy of the NV center spin, which requires that bias fields are applied along the anisotropy axis \cite{epstein05}. % to avoid energy level mixing and quenching of the optical spin contrast.

\begin{figure}
    \includegraphics[width=\linewidth]{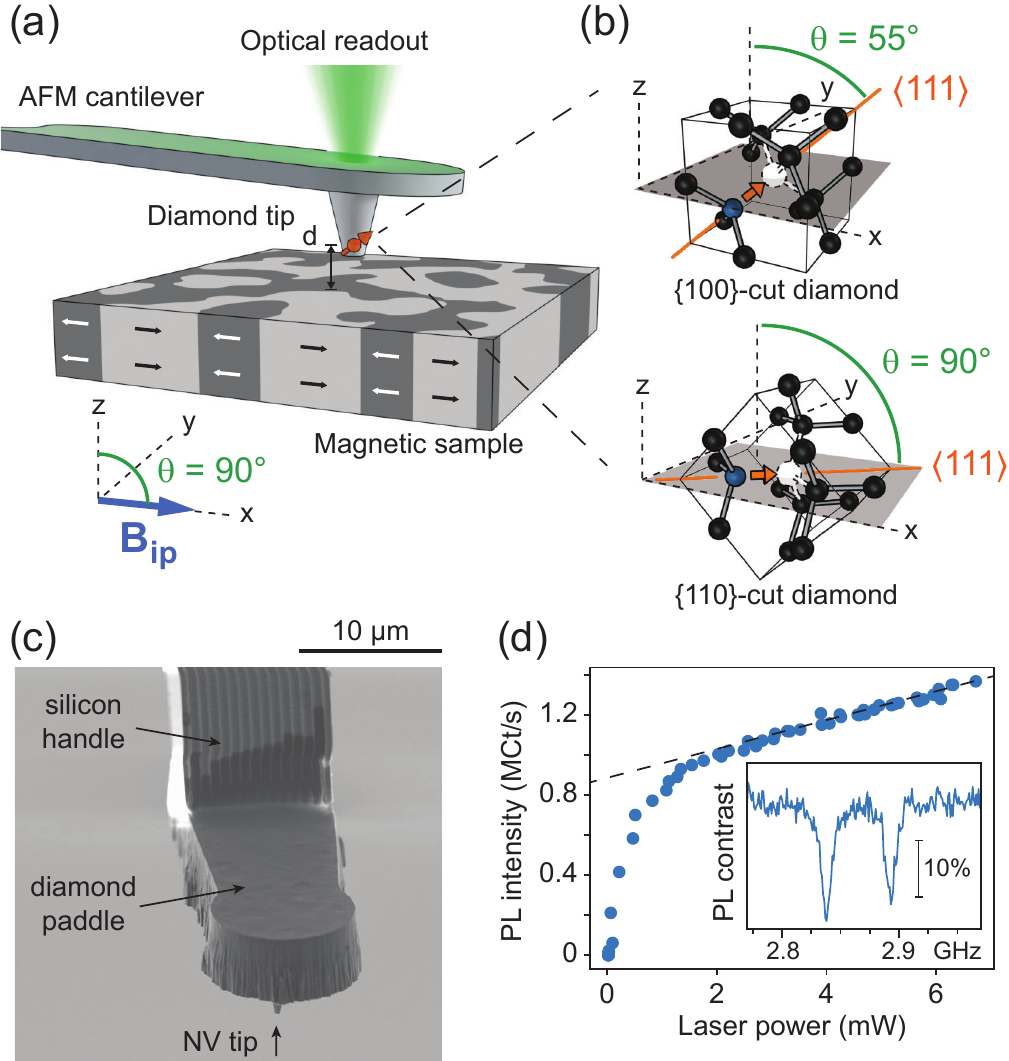}
    \caption{(a) Experimental arrangement.  A diamond scanning tip with an NV center at its apex is scanned over the magnetic surface of interest.  Optically-detected magnetic resonance is used to monitor shifts in the spin transition frequencies $\wpm$ due to the stray field of the sample.  In this work, we consider samples that require a variable in-plane magnetic field $\Bip$.
		(b) Crystallographic orientations of NV centers in diamond probes made from \{100\}-cut and \{110\}-cut crystals, respectively.  $\theta$ is the angle between the surface normal (the $z$ axis) and the NV anisotropy axis ($\langle 111\rangle$).  Only \{100\} probes can support $\theta=90^\circ$.
		(c) Scanning electron micrograph of a finished diamond probe including NV tip, diamond paddle and silicon handle.
		(d) Photo-luminescence (PL) saturation curve ($\Isat \sim 870\unit{kCt/s}$) and ODMR spectrum (inset, horizontal axis is microwave frequency) at $\Bip=1\unit{mT}$ for a representative \{110\} probe.
		}
	\label{fig1}
\end{figure}

In this work, we discuss fabrication and operation of diamond scanning probes with an in-plane spin anisotropy axis.  We reach the in-plane orientation by fabricating tips from $\{110\}$-cut diamond single crystals.  We demonstrate the performance of $\{110\}$-probes by imaging the stray field pattern from ferromagnetic domains of Co-NiO films at in-plane fields of up to about 40\,mT.  By comparison, we find that conventional, $\{100\}$-cut probes start loosing their sensitivity around 15\,mT and are completely quenched above 30\,mT.  Our work extends the applicability of scanning NV magnetometry to the wide range of magnetic structures that require large and variable in-plane fields for their investigation.

\vspace{0.5cm}
\textit{Theory -- }
Atomic magnetometry with NV centers relies on monitoring shifts in the spin transitions using optically-detected magnetic resonance (ODMR) spectroscopy \cite{budker07,schirhagl14}.  In its electronic ground state, the $S=1$ spin system presents two allowed transitions between the $|m_s=0\rangle$ and $|m_s=\pm 1\rangle$ spin sub-levels.  We label the frequencies of these transitions by $\wpm$.  At zero magnetic field, the $|\pm 1\rangle$ levels are degenerate and separated in energy by $D=2\pi\times2.87\unit{GHz}$ from the $|0\rangle$ level due to magneto-crystalline anisotropy \cite{schirhagl14}.  The anisotropy introduces a preferred quantization axis that coincides with the NV symmetry axis and lies along one of the four diagonals of the diamond unit cell ($\langle 111\rangle$ axes, see Fig.~\ref{fig1}).

Upon application of a magnetic field, the degeneracy is lifted and the $|+1\rangle$ ($|-1\rangle$) level shifts to higher (lower) energy.  The relevant spin Hamiltonian is given by \cite{schirhagl14}
\begin{align}
\mathcal{\hat H}/\hbar = D(\Sz^2-2/3) + \ye(\Bx\Sx + \By\Sy) + \ye \Bz\Sz ,
\end{align}
where $\ye = 2\pi\times 28\unit{GHz/T}$ is the electron gyromagnetic ratio and $\vecB = (\Bx,\By,\Bz)$ is the applied magnetic field. $\Sx$, $\Sy$, and $\Sz$ are spin-1 operators.  The coordinates $(\bar{x},\bar{y},\bar{z})$ of the NV frame of reference are tilted with respect to the laboratory frame by
$(\bar{x}=x\cos\theta+z\sin\theta,\bar{y},\bar{z}=z\cos\theta-x\sin\theta)$, see Fig.~\ref{fig1}(b), such that the $\bar z$ axis coincides with the $\langle 111\rangle$ anisotropy axis.

Because of the magneto-crystalline anisotropy, fields applied along the quantization axis ($\Bz$) have different effects on the energy levels from those applied transverse to the quantization axis ($\Bx, \By$).  In particular, parallel fields $\Bz$ commute with the magneto-crystalline anisotropy and lead to a linear Zeeman shift
\begin{equation}
\wpm = D \pm \ye\Bz \ .
\label{eq:linear}
\end{equation}
By contrast, a transverse field in the $\bar x$ (or $\bar y$) direction leads to a mixing between energy levels and a quadratic shift in the transition frequencies,
\begin{equation}
\wpm \approx D \pm \ye\Bz + \frac{3\ye^2\Bx^2}{2D}
\label{eq:quadratic}
\end{equation}
where the approximation is best for $\Bz,\Bx\ll D$ and $\Bz$ not much smaller than $\Bx$.

The spin mixing due to off-axis fields has a dramatic effect on the photo-luminescence (PL) intensity.  Because the optical transitions are no longer spin conserving, the decay rates from the excited state become similar for both spin states.  As a consequence, the $\sim$30\% higher PL intensity of $|0\rangle$ compared to the $|\pm 1\rangle$ spin state vanishes, and spin-state readout is no longer possible.  This prohibits magnetometry operation.  In practice, the effects of spin mixing are complex \cite{dreau11}, and transverse fields as small as a few mT already lead to a loss in PL contrast.

Since scanning magnetometer probes are cut from single-crystal diamond, the crystal orientation and preferred field direction are locked to the laboratory coordinate system.  For standard diamond probes fabricated from crystals with an exposed \{100\} facet, the preferred field direction is at an angle of $\theta = 55^\circ$ from the surface normal (see Fig. \ref{fig1}(b)).  This is an odd angle for applications to magnetic materials, where either out-of-plane (OOP, $\theta=0^\circ$) or in-plane (IP, $\theta=90^\circ$) bias fields are typically required. %, for example, for studying magnetic hysteresis. 

The obvious route towards OOP or IP magnetometry is the use of other crystal facets that support the desired NV orientations.  In particular, \{111\}-cut diamond exhibits a $\theta=0^\circ$ orientation, while \{110\}-cut diamond contains two $\theta=90^\circ$ orientations.  Fabrication of different crystal facets with high surface quality, however, is highly challenging due to diamond's superb hardness and difficult growth.  Recently, engineering \{111\} crystals with large concentrations of vertically-aligned NV \cite{neu14} and realization of scanning diamond probes from \{111\} diamond \cite{rohner19} has been reported.  In this work, we describe fabrication of IP-oriented diamond probes and demonstrate scanning imaging in large IP bias fields with little loss in optical PL contrast.

\vspace{0.5cm}
\textit{Probe fabrication -- }
The starting material for our diamond probes are high-purity single-crystal diamond plates with a main \{110\} facet grown by high-pressure-high-temperature synthesis (NDT LLC, Russia).  The intrinsic nitrogen concentration is not measured, but is expected to be less than 5\,ppb as no native NV centers can be detected in the raw material.  We form NV centers by ion implantation ($^{15}$N$^{+}$, 7\,keV, $3\ee{10}$ ions per cm$^2$, CuttingEdgeIons Inc.) and vacuum annealing (880°C, $p<5\ee{-8}\unit{mbar}$, 2\,h).
To fabricate the tips, we use a series of e-beam lithography and inductively-coupled plasma (ICP, Oxford Instruments PlasmaPro 100) etching steps.
Fig.~\ref{fig1}(c) shows an SEM micrograph of a finished tip.  PL characterization shows saturation count rates between $\Isat = 800-1200\unit{kCt/s}$ and spin contrast between $\eps = 20-25\%$, see Fig.~\ref{fig1}(d). These PL count rate and contrast values are comparable to those observed on \{100\} tips of similar geometry \cite{qzabre}.
Two crystals are processed in this study.  With the first batch, we observe a reduced photo-stability of the NV centers ($\sim 50\%$ of NV centers bleached within hours of use), likely due to charge conversion \cite{rondin10}.  NV centers from the second batch do not show this issue.  Overall, 14 probes from the first batch and 2 probes from the second batch are used in this study. (Several additional probes from the second crystal have been used in other measurements.)

\vspace{0.5cm}
\textit{Device and Setup -- }
Our scanning diamond probes are assembled by gluing the diamond paddle with the scanning tip (Fig.~\ref{fig1}(c)) to a silicon handle structure, which itself is attached to a quartz tuning fork. The piezoelectric tuning fork serves as an AFM position feedback.  For initial characterization, we use a commercial magnetometer setup (QSM, QZabre AG) equipped with a vector electromagnet to apply a programmable magnetic field.  For scanning experiments, we mount the assembled probe chip in a custom-built scanning diamond magnetometer \cite{wornle21thesis} and apply the in-plane magnetic field by placing two small permanent magnets on either side of the sample (Fig.~S1~\cite{supplementary}).  In both setups, we measure the ODMR resonance by continuous illumination with $520\unit{nm}$ laser light and confocal collection of the emitted luminescent light with a single photon detector. The microwave drive signal is applied by guiding a microwave current through a thin wire suspended near the sample.  All measurements are conducted in ambient conditions.
\begin{figure}
    \includegraphics[width=\linewidth]{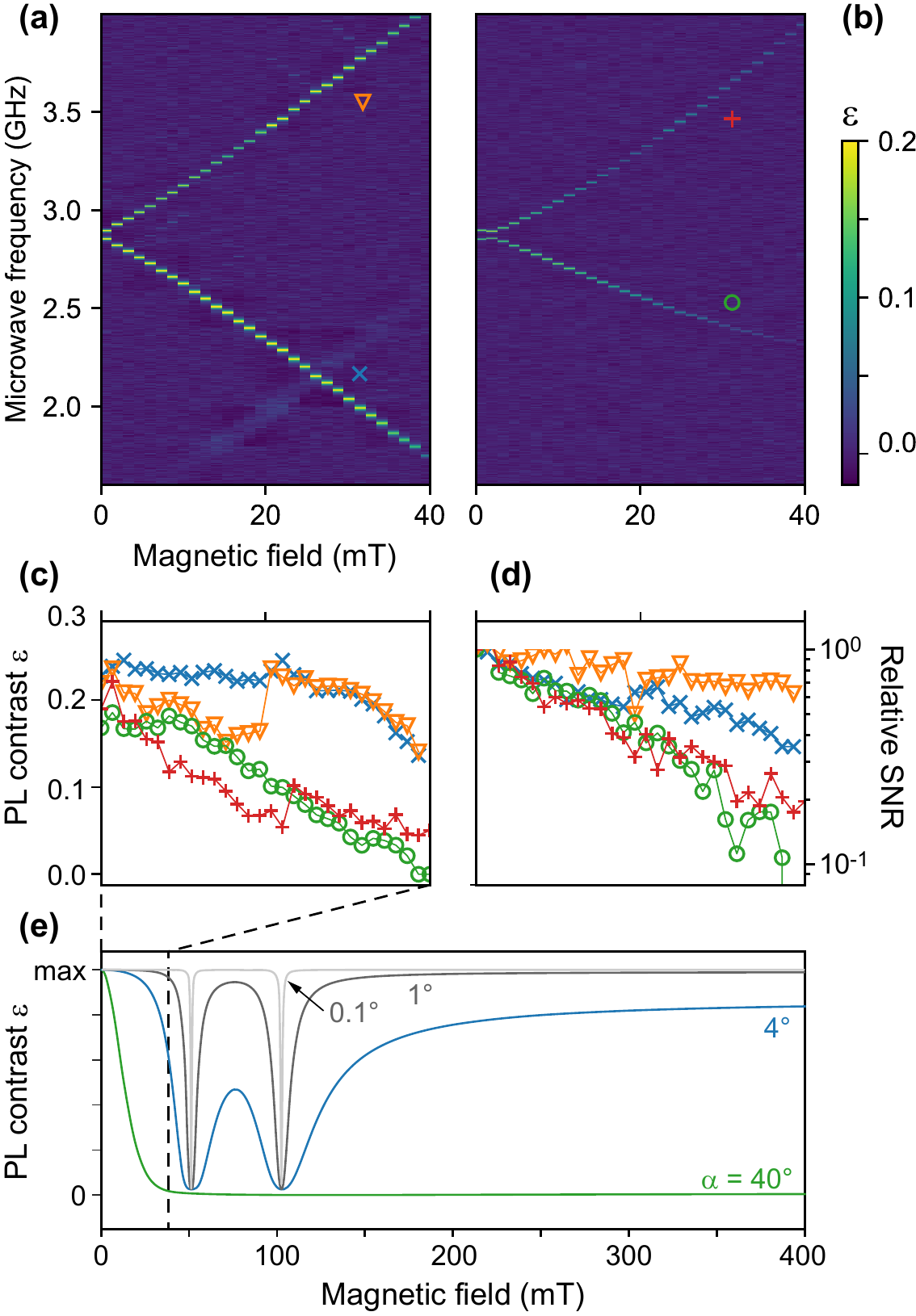}
    \caption{(a,b) Series of ODMR spectra for an increasing in-plane bias field $\Bip$.
		Color code is PL contrast and symbols identify $\wa$ ($\triangledown$) and $\wb$ ($\times$) for the $\{110\}$ probe, and $\wa$ ($+$) and $\wb$ ($\circ$) for the $\{100\}$ probe.  For the $\{100\}$ probe, the misalignment between the NV center axis and the bias field axis is approximately $\alpha \approx 40^\circ$.
		(c) PL contrast $\eps$ as a function of $\Bip$.
		(d) Relative $\SNR$ (see Eq.~(\ref{eq:snr})) as a function of $\Bip$.
		The reduced contrast and anomaly near $20\unit{mT}$ for ($\triangledown$,$+$) is due to a resonance in the microwave feed-line, causing a variation in microwave power and an associated variation in $\eps$ and linewidth $\Dw$ \cite{dreau11}. Note that the $\SNR$ is mostly unaffected.
    (e) Expected contrast vs.~field for various misalignment angles $\alpha$, based on the model from Ref.~\cite{tetienne12}.
		}
    \label{fig2}
\end{figure}
\begin{figure*}
    \includegraphics[width=0.77\linewidth]{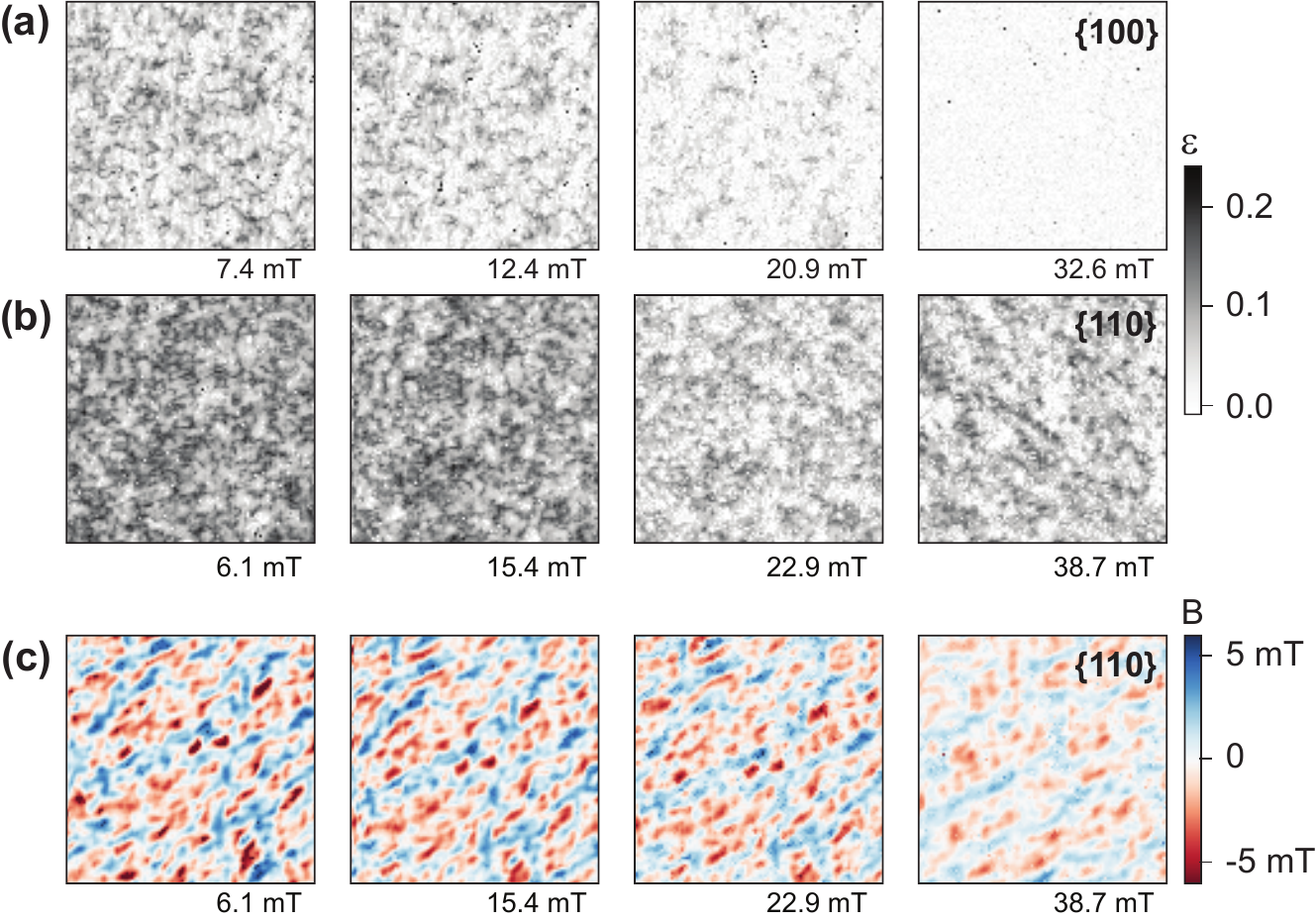}
    \caption{Magnetometry scans of a Co-NiO film for increasing magnetic in-plane fields $\Bip = 0-40\unit{mT}$.
		(a) Map of the PL contrast $\eps$ recorded with a conventional \{100\} probe.
		(b) Map of the PL contrast $\eps$ recorded with an in-plane \{110\} probe.
		(c) Magnetic field map $B$ recorded using the in-plane \{110\} probe.
		Note that the scan for $\Bip=38.7\unit{mT}$ was recorded in a different region on the sample.
		All scans are $3\um$ wide.
    }
    \label{fig3}
\end{figure*}

\vspace{0.5cm}
\textit{Experimental results -- }
In a first set of experiments, we analyze the PL contrast as a function of applied bias field $\Bip$.  Fig.~\ref{fig2}(a) shows a series of ODMR spectra for the \{110\} tip plotted in a two-dimensional PL map.  Here, the bias field is aligned (to within a few degrees) with the NV symmetry axis.  As expected, the response of the resonance frequencies is linear in $\Bip$ (Eq.~(\ref{eq:linear})) and the PL contrast remains high.
By contrast, for the conventional \{100\} tip, the bias is misaligned by approximately $\alpha=40^\circ$ from the NV axis.  Now, the dispersion is markedly non-linear (Eq.~(\ref{eq:quadratic})) and the PL contrast strongly reduced once $\Bip\gtrsim 15\unit{mT}$.
Figs.~\ref{fig2}(c,d) provide further analysis by plotting the PL contrast $\eps$ and relative signal-to-noise ratio (SNR) as a function of $\Bip$.  We extract quantities by fitting spectra with a Lorentzian \cite{dreau11}, yielding, besides $\wpm$ and $\eps$, estimates for the resonance linewidth $\Dw$ (full width at half maximum) and the steady-state PL count rate $I_0$.
From these values, we compute the minimum detectable field \cite{dreau11,schirhagl14},
\begin{align}
\Bmin = \frac{\Dw}{2\ye\eps\sqrt{I_0}} \ ,
\label{eq:bmin}
\end{align}
and the relative SNR:
\begin{align}
\SNR(\Bip) := \frac{\Bmin^{(\Bip=0)}} {\Bmin^{(\Bip)}} \ .
\label{eq:snr}
\end{align}
Figs.~\ref{fig2}(c,d) plot $\eps$ and the $\SNR$ for both types of probes.  Clearly, for the conventional $\{100\}$ tips, both $\eps$ and the $\SNR$ rapidly deteriorate as the bias field exceeds $\sim 15\unit{mT}$, while the \{110\} tips maintain almost full PL contrast and SNR beyond $\Bip=30\unit{mT}$.  At even higher fields, also the \{110\} tips show a reduction in PL contrast due to a residual field misalignment ($\sim 4^\circ$).  Fig.~\ref{fig2}(e) provides a simulation of $\eps$ up to $400\unit{mT}$ based on a rate model of the optical excitation \cite{tetienne12}, confirming that the PL contrast is extremely sensitive to the field misalignment.  Fig.~\ref{fig2}(e) also shows that for a proper field alignment, full PL contrast can in principle be maintained for the entire field range, except for values close to the spin zero-crossings near $50\unit{mT}$ and $100\unit{mT}$.

\vspace{0.5cm}
\textit{Imaging -- }
To demonstrate nanoscale magnetic imaging, we perform scanning magnetometry of the domain pattern formed in a 2.5-nm-thin Co film sandwiched between two NiO layers.  Such a sandwich structure of ferromagnetic (Co) and antiferromagnetic (NiO) films has considerable potential for applications in spintronics \cite{mazalski20}. The sample is fabricated by magnetron sputtering on a Si substrate, with the following stack-up: Si/SiO$_2$(400)/Ta(3)/Pt(5)/NiO(5)/Co(2.5)/NiO(5)/Pt(3), where thicknesses are in nm.  The NiO film in this sample acts as a pinning layer for the Co moments and increases the latter's coercivity via exchange bias.  Fig.~S2 shows a hysteresis loop of the full film indicating that the in-plane coercivity is around $40-50\unit{mT}$ (Fig.~S2~\cite{supplementary}).

Figs.~\ref{fig3}(a,b) show maps of the PL contrast obtained by scanning the respective tips over the sample surface.  The in-plane bias field increases from left to right.  The PL contrast maps show a pattern of dark (white) regions, indicating areas where the NV centers experience a small (large) transverse field.  The contrast maps therefore loosely reflect the Co-NiO domain structure \cite{wornle19}.  Importantly, as $\Bip$ increases, the PL contrast for \{100\} probes turns to complete white, \ie, a total loss of PL contrast.  Hence, the probe entirely looses its ability to sense the stray field above about~$20\unit{mT}$ and magnetometry analysis is no longer possible.  By contrast, for the \{110\} probe, almost full PL contrast is maintained up to maximum accessible field $\Bip \approx 40\unit{mT}$.  The persistent PL contrast is in agreement with the data of Fig.~\ref{fig2}(c).
%Similar to Fig.~\ref{fig2}c, some PL contrast is slightly degraded because of imperfect alignment of the magnetic field.

In Fig.~\ref{fig3}(c), we use the \{110\} probe to perform magnetometry of the Co-NiO sample over the full $\Bip$ range.  Each panel plots the in-plane component of the sample's stray field $B = (\wa-\wa^0)/\ye$, where $\wa^0$ is the resonance frequency measured after retracting the tip by a few $\um$ from the sample surface.  We observe magnetic structures with a typical length scale of $100-200\unit{nm}$, which is a reasonable number for the average domain size for the Co film \cite{wu09,mazalski20}.  While the magnetic pattern is essentially unchanged for the three panels $\Bip<30\unit{mT}$, a clear reduction in the stray field is observed in the last panel where the applied field is close to the coercive field.  Note that this reduction is unaffected by possible variations in the PL contrast because the latter only influences the sensitivity, but not the measured stray field value.  In addition, there is an apparent slight increase in the domain size.  For even larger bias fields, complete saturation of the sample would be expected; however, these fields are not accessible with our present experimental arrangement.  Overall, Fig.~\ref{fig3} clearly demonstrates our ability to study stray field patterns in large in-plane fields without a significant loss in sensitivity.

\vspace{0.5cm}
\textit{Outlook -- }
In summary, we demonstrate the fabrication and application of scanning probes made from \{110\}-cut diamond. These probes can host NV centers with their anisotropy axis parallel to the sample surface. This arrangement makes it possible to apply a in-plane field magnetic field to samples without loosing the NV center's sensing capability.  As an example, we demonstrate quantitative scanning magnetometry of a Co-NiO film in applied fields of up to $\sim 40\unit{mT}$.  Looking forward, the ability of applying purely in-plane bias fields is important to a number of applications in the fields of spintronics and thin-film magnetism, for example, to break the symmetry and allow current-induced switching of ferromagnetic layers \cite{miron11}.

%%%%%%%%%%%%%% Acknowledgments

\vspace{0.5cm}
\textit{Acknowledgments -- }
The authors thank Simon Josephy for assistance in the operation of the vector magnet and scanning microscope and Patrick Scheidegger for helpful comments on the manuscript.
This work was supported by Project Grant No. 200020\_175600 and PZ00P2\_179944 of the Swiss National Science Foundation (SNSF), by the National Center of Competence in Research in Quantum Science and Technology (NCCR QSIT) of the SNSF, by Innosuisse grant 43106.1 IP-ENG, and by the Advancing Science and TEchnology thRough dIamond Quantum Sensing (ASTERIQS) program, Grant No. 820394, of the European Commission.

%%%%%%%%%%%%% Statements

\vspace{0.5cm}
\textit{Conflict of interest -- }
The authors have no conflicts to disclose.

\vspace{0.5cm}
\textit{Data availability statement -- }
The data that support the findings of this study are available from the corresponding author upon reasonable request.

\bibliography{library}

\begin{thebibliography}{39}%
\makeatletter
\providecommand \@ifxundefined [1]{%
 \@ifx{#1\undefined}
}%
\providecommand \@ifnum [1]{%
 \ifnum #1\expandafter \@firstoftwo
 \else \expandafter \@secondoftwo
 \fi
}%
\providecommand \@ifx [1]{%
 \ifx #1\expandafter \@firstoftwo
 \else \expandafter \@secondoftwo
 \fi
}%
\providecommand \natexlab [1]{#1}%
\providecommand \enquote  [1]{``#1''}%
\providecommand \bibnamefont  [1]{#1}%
\providecommand \bibfnamefont [1]{#1}%
\providecommand \citenamefont [1]{#1}%
\providecommand \href@noop [0]{\@secondoftwo}%
\providecommand \href [0]{\begingroup \@sanitize@url \@href}%
\providecommand \@href[1]{\@@startlink{#1}\@@href}%
\providecommand \@@href[1]{\endgroup#1\@@endlink}%
\providecommand \@sanitize@url [0]{\catcode `\\12\catcode `\$12\catcode
  `\&12\catcode `\#12\catcode `\^12\catcode `\_12\catcode `\%12\relax}%
\providecommand \@@startlink[1]{}%
\providecommand \@@endlink[0]{}%
\providecommand \url  [0]{\begingroup\@sanitize@url \@url }%
\providecommand \@url [1]{\endgroup\@href {#1}{\urlprefix }}%
\providecommand \urlprefix  [0]{URL }%
\providecommand \Eprint [0]{\href }%
\providecommand \doibase [0]{https://doi.org/}%
\providecommand \selectlanguage [0]{\@gobble}%
\providecommand \bibinfo  [0]{\@secondoftwo}%
\providecommand \bibfield  [0]{\@secondoftwo}%
\providecommand \translation [1]{[#1]}%
\providecommand \BibitemOpen [0]{}%
\providecommand \bibitemStop [0]{}%
\providecommand \bibitemNoStop [0]{.\EOS\space}%
\providecommand \EOS [0]{\spacefactor3000\relax}%
\providecommand \BibitemShut  [1]{\csname bibitem#1\endcsname}%
\let\auto@bib@innerbib\@empty
%</preamble>
\bibitem [{\citenamefont {Kosub}\ \emph {et~al.}(2017)\citenamefont {Kosub},
  \citenamefont {Kopte}, \citenamefont {Huhne}, \citenamefont {Appel},
  \citenamefont {Shields}, \citenamefont {Maletinsky}, \citenamefont {Hubner},
  \citenamefont {Liedke}, \citenamefont {Fassbender}, \citenamefont {Schmidt},\
  and\ \citenamefont {Makarov}}]{kosub17}%
  \BibitemOpen
  \bibfield  {author} {\bibinfo {author} {\bibfnamefont {T.}~\bibnamefont
  {Kosub}}, \bibinfo {author} {\bibfnamefont {M.}~\bibnamefont {Kopte}},
  \bibinfo {author} {\bibfnamefont {R.}~\bibnamefont {Huhne}}, \bibinfo
  {author} {\bibfnamefont {P.}~\bibnamefont {Appel}}, \bibinfo {author}
  {\bibfnamefont {B.}~\bibnamefont {Shields}}, \bibinfo {author} {\bibfnamefont
  {P.}~\bibnamefont {Maletinsky}}, \bibinfo {author} {\bibfnamefont
  {R.}~\bibnamefont {Hubner}}, \bibinfo {author} {\bibfnamefont {M.~O.}\
  \bibnamefont {Liedke}}, \bibinfo {author} {\bibfnamefont {J.}~\bibnamefont
  {Fassbender}}, \bibinfo {author} {\bibfnamefont {O.~G.}\ \bibnamefont
  {Schmidt}},\ and\ \bibinfo {author} {\bibfnamefont {D.}~\bibnamefont
  {Makarov}},\ }\bibfield  {title} {\bibinfo {title} {Purely antiferromagnetic
  magnetoelectric random access memory},\ }\href
  {https://doi.org/10.1038/ncomms13985} {\bibfield  {journal} {\bibinfo
  {journal} {Nature Communications}\ }\textbf {\bibinfo {volume} {8}},\
  \bibinfo {pages} {13985} (\bibinfo {year} {2017})}\BibitemShut {NoStop}%
\bibitem [{\citenamefont {Appel}\ \emph {et~al.}(2019)\citenamefont {Appel},
  \citenamefont {Shields}, \citenamefont {Kosub}, \citenamefont {Hedrich},
  \citenamefont {Hubner}, \citenamefont {Fassbender}, \citenamefont {Makarov},\
  and\ \citenamefont {Maletinsky}}]{appel19}%
  \BibitemOpen
  \bibfield  {author} {\bibinfo {author} {\bibfnamefont {P.}~\bibnamefont
  {Appel}}, \bibinfo {author} {\bibfnamefont {B.~J.}\ \bibnamefont {Shields}},
  \bibinfo {author} {\bibfnamefont {T.}~\bibnamefont {Kosub}}, \bibinfo
  {author} {\bibfnamefont {N.}~\bibnamefont {Hedrich}}, \bibinfo {author}
  {\bibfnamefont {R.}~\bibnamefont {Hubner}}, \bibinfo {author} {\bibfnamefont
  {J.}~\bibnamefont {Fassbender}}, \bibinfo {author} {\bibfnamefont
  {D.}~\bibnamefont {Makarov}},\ and\ \bibinfo {author} {\bibfnamefont
  {P.}~\bibnamefont {Maletinsky}},\ }\bibfield  {title} {\bibinfo {title}
  {Nanomagnetism of magnetoelectric granular thin-film antiferromagnets},\
  }\href {https://doi.org/10.1021/acs.nanolett.8b04681} {\bibfield  {journal}
  {\bibinfo  {journal} {Nano Lett.}\ }\textbf {\bibinfo {volume} {19}},\
  \bibinfo {pages} {1682} (\bibinfo {year} {2019})}\BibitemShut {NoStop}%
\bibitem [{\citenamefont {Wornle}\ \emph {et~al.}(2019)\citenamefont {Wornle},
  \citenamefont {Welter}, \citenamefont {Kaspar}, \citenamefont {Olejnik},
  \citenamefont {Novak}, \citenamefont {Campion}, \citenamefont {Wadley},
  \citenamefont {Jungwirth}, \citenamefont {Degen},\ and\ \citenamefont
  {Gambardella}}]{wornle19}%
  \BibitemOpen
  \bibfield  {author} {\bibinfo {author} {\bibfnamefont {M.~S.}\ \bibnamefont
  {Wornle}}, \bibinfo {author} {\bibfnamefont {P.}~\bibnamefont {Welter}},
  \bibinfo {author} {\bibfnamefont {Z.}~\bibnamefont {Kaspar}}, \bibinfo
  {author} {\bibfnamefont {K.}~\bibnamefont {Olejnik}}, \bibinfo {author}
  {\bibfnamefont {V.}~\bibnamefont {Novak}}, \bibinfo {author} {\bibfnamefont
  {R.~P.}\ \bibnamefont {Campion}}, \bibinfo {author} {\bibfnamefont
  {P.}~\bibnamefont {Wadley}}, \bibinfo {author} {\bibfnamefont
  {T.}~\bibnamefont {Jungwirth}}, \bibinfo {author} {\bibfnamefont {C.~L.}\
  \bibnamefont {Degen}},\ and\ \bibinfo {author} {\bibfnamefont
  {P.}~\bibnamefont {Gambardella}},\ }\bibfield  {title} {\bibinfo {title}
  {Current-induced fragmentation of antiferromagnetic domains},\ }\href
  {https://arxiv.org/abs/1912.05287} {\bibfield  {journal} {\bibinfo  {journal}
  {arXiv:1912.05287}\ } (\bibinfo {year} {2019})}\BibitemShut {NoStop}%
\bibitem [{\citenamefont {Dovzhenko}\ \emph {et~al.}(2018)\citenamefont
  {Dovzhenko}, \citenamefont {Casola}, \citenamefont {Schlotter}, \citenamefont
  {Zhou}, \citenamefont {Buttner}, \citenamefont {Walsworth}, \citenamefont
  {Beach},\ and\ \citenamefont {Yacoby}}]{dovzhenko18}%
  \BibitemOpen
  \bibfield  {author} {\bibinfo {author} {\bibfnamefont {Y.}~\bibnamefont
  {Dovzhenko}}, \bibinfo {author} {\bibfnamefont {F.}~\bibnamefont {Casola}},
  \bibinfo {author} {\bibfnamefont {S.}~\bibnamefont {Schlotter}}, \bibinfo
  {author} {\bibfnamefont {T.~X.}\ \bibnamefont {Zhou}}, \bibinfo {author}
  {\bibfnamefont {F.}~\bibnamefont {Buttner}}, \bibinfo {author} {\bibfnamefont
  {R.~L.}\ \bibnamefont {Walsworth}}, \bibinfo {author} {\bibfnamefont
  {G.~S.~D.}\ \bibnamefont {Beach}},\ and\ \bibinfo {author} {\bibfnamefont
  {A.}~\bibnamefont {Yacoby}},\ }\bibfield  {title} {\bibinfo {title}
  {Magnetostatic twists in room-temperature skyrmions explored by
  nitrogen-vacancy center spin texture reconstruction},\ }\href
  {https://doi.org/10.1038/s41467-018-05158-9} {\bibfield  {journal} {\bibinfo
  {journal} {Nature Communications}\ }\textbf {\bibinfo {volume} {9}},\
  \bibinfo {pages} {2712} (\bibinfo {year} {2018})}\BibitemShut {NoStop}%
\bibitem [{\citenamefont {Gross}\ \emph {et~al.}(2017)\citenamefont {Gross},
  \citenamefont {Akhtar}, \citenamefont {Garcia}, \citenamefont {Martinez},
  \citenamefont {Chouaieb}, \citenamefont {Garcia}, \citenamefont {Carretero},
  \citenamefont {Arthelemy}, \citenamefont {Appel}, \citenamefont {Maletinsky},
  \citenamefont {Kim}, \citenamefont {Chauleau}, \citenamefont {Jaouen},
  \citenamefont {Viret}, \citenamefont {Bibes}, \citenamefont {Fusil},\ and\
  \citenamefont {Jacques}}]{gross17}%
  \BibitemOpen
  \bibfield  {author} {\bibinfo {author} {\bibfnamefont {I.}~\bibnamefont
  {Gross}}, \bibinfo {author} {\bibfnamefont {W.}~\bibnamefont {Akhtar}},
  \bibinfo {author} {\bibfnamefont {V.}~\bibnamefont {Garcia}}, \bibinfo
  {author} {\bibfnamefont {L.~J.}\ \bibnamefont {Martinez}}, \bibinfo {author}
  {\bibfnamefont {S.}~\bibnamefont {Chouaieb}}, \bibinfo {author}
  {\bibfnamefont {K.}~\bibnamefont {Garcia}}, \bibinfo {author} {\bibfnamefont
  {C.}~\bibnamefont {Carretero}}, \bibinfo {author} {\bibfnamefont
  {B.}~\bibnamefont {Arthelemy}}, \bibinfo {author} {\bibfnamefont
  {P.}~\bibnamefont {Appel}}, \bibinfo {author} {\bibfnamefont
  {P.}~\bibnamefont {Maletinsky}}, \bibinfo {author} {\bibfnamefont {J.~V.}\
  \bibnamefont {Kim}}, \bibinfo {author} {\bibfnamefont {J.~Y.}\ \bibnamefont
  {Chauleau}}, \bibinfo {author} {\bibfnamefont {N.}~\bibnamefont {Jaouen}},
  \bibinfo {author} {\bibfnamefont {M.}~\bibnamefont {Viret}}, \bibinfo
  {author} {\bibfnamefont {M.}~\bibnamefont {Bibes}}, \bibinfo {author}
  {\bibfnamefont {S.}~\bibnamefont {Fusil}},\ and\ \bibinfo {author}
  {\bibfnamefont {V.}~\bibnamefont {Jacques}},\ }\bibfield  {title} {\bibinfo
  {title} {Real-space imaging of non-collinear antiferromagnetic order with a
  single-spin magnetometer},\ }\href {https://doi.org/10.1038/nature23656}
  {\bibfield  {journal} {\bibinfo  {journal} {Nature}\ }\textbf {\bibinfo
  {volume} {549}},\ \bibinfo {pages} {252} (\bibinfo {year}
  {2017})}\BibitemShut {NoStop}%
\bibitem [{\citenamefont {Ariyaratne}\ \emph {et~al.}(2018)\citenamefont
  {Ariyaratne}, \citenamefont {Bluvstein}, \citenamefont {Myers},\ and\
  \citenamefont {Jayich}}]{ariyaratne18}%
  \BibitemOpen
  \bibfield  {author} {\bibinfo {author} {\bibfnamefont {A.}~\bibnamefont
  {Ariyaratne}}, \bibinfo {author} {\bibfnamefont {D.}~\bibnamefont
  {Bluvstein}}, \bibinfo {author} {\bibfnamefont {B.~A.}\ \bibnamefont
  {Myers}},\ and\ \bibinfo {author} {\bibfnamefont {A.~C.~B.}\ \bibnamefont
  {Jayich}},\ }\bibfield  {title} {\bibinfo {title} {Nanoscale electrical
  conductivity imaging using a nitrogen-vacancy center in diamond},\ }\href
  {https://doi.org/10.1038/s41467-018-04798-1} {\bibfield  {journal} {\bibinfo
  {journal} {Nature Communications}\ }\textbf {\bibinfo {volume} {9}},\
  \bibinfo {pages} {2406} (\bibinfo {year} {2018})}\BibitemShut {NoStop}%
\bibitem [{\citenamefont {Thiel}\ \emph {et~al.}(2016)\citenamefont {Thiel},
  \citenamefont {Rohner}, \citenamefont {Ganzhorn}, \citenamefont {Appel},
  \citenamefont {Neu}, \citenamefont {Muller}, \citenamefont {Kleiner},
  \citenamefont {Koelle},\ and\ \citenamefont {Maletinsky}}]{thiel16}%
  \BibitemOpen
  \bibfield  {author} {\bibinfo {author} {\bibfnamefont {L.}~\bibnamefont
  {Thiel}}, \bibinfo {author} {\bibfnamefont {D.}~\bibnamefont {Rohner}},
  \bibinfo {author} {\bibfnamefont {M.}~\bibnamefont {Ganzhorn}}, \bibinfo
  {author} {\bibfnamefont {P.}~\bibnamefont {Appel}}, \bibinfo {author}
  {\bibfnamefont {E.}~\bibnamefont {Neu}}, \bibinfo {author} {\bibfnamefont
  {B.}~\bibnamefont {Muller}}, \bibinfo {author} {\bibfnamefont
  {R.}~\bibnamefont {Kleiner}}, \bibinfo {author} {\bibfnamefont
  {D.}~\bibnamefont {Koelle}},\ and\ \bibinfo {author} {\bibfnamefont
  {P.}~\bibnamefont {Maletinsky}},\ }\bibfield  {title} {\bibinfo {title}
  {Quantitative nanoscale vortex imaging using a cryogenic quantum
  magnetometer},\ }\href {https://doi.org/10.1038/NNANO.2016.63} {\bibfield
  {journal} {\bibinfo  {journal} {Nat. Nanotechnol.}\ }\textbf {\bibinfo
  {volume} {11}},\ \bibinfo {pages} {677} (\bibinfo {year} {2016})}\BibitemShut
  {NoStop}%
\bibitem [{\citenamefont {Nowack}\ \emph {et~al.}(2013)\citenamefont {Nowack},
  \citenamefont {Spanton}, \citenamefont {Baenninger}, \citenamefont {Konig},
  \citenamefont {Kirtley}, \citenamefont {Kalisky}, \citenamefont {Ames},
  \citenamefont {Leubner}, \citenamefont {Brune}, \citenamefont {Buhmann},
  \citenamefont {Molenkamp}, \citenamefont {Goldhaber-Gordon},\ and\
  \citenamefont {Moler}}]{nowack13}%
  \BibitemOpen
  \bibfield  {author} {\bibinfo {author} {\bibfnamefont {K.~C.}\ \bibnamefont
  {Nowack}}, \bibinfo {author} {\bibfnamefont {E.~M.}\ \bibnamefont {Spanton}},
  \bibinfo {author} {\bibfnamefont {M.}~\bibnamefont {Baenninger}}, \bibinfo
  {author} {\bibfnamefont {M.}~\bibnamefont {Konig}}, \bibinfo {author}
  {\bibfnamefont {J.~R.}\ \bibnamefont {Kirtley}}, \bibinfo {author}
  {\bibfnamefont {B.}~\bibnamefont {Kalisky}}, \bibinfo {author} {\bibfnamefont
  {C.}~\bibnamefont {Ames}}, \bibinfo {author} {\bibfnamefont {P.}~\bibnamefont
  {Leubner}}, \bibinfo {author} {\bibfnamefont {C.}~\bibnamefont {Brune}},
  \bibinfo {author} {\bibfnamefont {H.}~\bibnamefont {Buhmann}}, \bibinfo
  {author} {\bibfnamefont {L.~W.}\ \bibnamefont {Molenkamp}}, \bibinfo {author}
  {\bibfnamefont {D.}~\bibnamefont {Goldhaber-Gordon}},\ and\ \bibinfo {author}
  {\bibfnamefont {K.~A.}\ \bibnamefont {Moler}},\ }\bibfield  {title} {\bibinfo
  {title} {Imaging currents in hgte quantum wells in the quantum spin hall
  regime},\ }\href {https://doi.org/10.1038/nmat3682} {\bibfield  {journal}
  {\bibinfo  {journal} {Nat. Mater.}\ }\textbf {\bibinfo {volume} {12}},\
  \bibinfo {pages} {787} (\bibinfo {year} {2013})}\BibitemShut {NoStop}%
\bibitem [{\citenamefont {Chang}\ \emph {et~al.}(2017)\citenamefont {Chang},
  \citenamefont {Eichler}, \citenamefont {Rhensius}, \citenamefont
  {Lorenzelli},\ and\ \citenamefont {Degen}}]{chang17}%
  \BibitemOpen
  \bibfield  {author} {\bibinfo {author} {\bibfnamefont {K.}~\bibnamefont
  {Chang}}, \bibinfo {author} {\bibfnamefont {A.}~\bibnamefont {Eichler}},
  \bibinfo {author} {\bibfnamefont {J.}~\bibnamefont {Rhensius}}, \bibinfo
  {author} {\bibfnamefont {L.}~\bibnamefont {Lorenzelli}},\ and\ \bibinfo
  {author} {\bibfnamefont {C.~L.}\ \bibnamefont {Degen}},\ }\bibfield  {title}
  {\bibinfo {title} {Nanoscale imaging of current density with a single-spin
  magnetometer},\ }\href {https://doi.org/10.1021/acs.nanolett.6b05304}
  {\bibfield  {journal} {\bibinfo  {journal} {Nano Letters}\ }\textbf {\bibinfo
  {volume} {17}},\ \bibinfo {pages} {2367} (\bibinfo {year}
  {2017})}\BibitemShut {NoStop}%
\bibitem [{\citenamefont {Ku}\ \emph {et~al.}(2020)\citenamefont {Ku},
  \citenamefont {Zhou}, \citenamefont {Li}, \citenamefont {Shin}, \citenamefont
  {Shi}, \citenamefont {Burch}, \citenamefont {Anderson}, \citenamefont
  {Pierce}, \citenamefont {Xie}, \citenamefont {Hamo}, \citenamefont {Vool},
  \citenamefont {Zhang}, \citenamefont {Casola}, \citenamefont {Taniguchi},
  \citenamefont {Watanabe}, \citenamefont {Fogler}, \citenamefont {Kim},
  \citenamefont {Yacoby},\ and\ \citenamefont {Walsworth}}]{ku20}%
  \BibitemOpen
  \bibfield  {author} {\bibinfo {author} {\bibfnamefont {M.~J.~H.}\
  \bibnamefont {Ku}}, \bibinfo {author} {\bibfnamefont {T.~X.}\ \bibnamefont
  {Zhou}}, \bibinfo {author} {\bibfnamefont {Q.}~\bibnamefont {Li}}, \bibinfo
  {author} {\bibfnamefont {Y.~J.}\ \bibnamefont {Shin}}, \bibinfo {author}
  {\bibfnamefont {J.~K.}\ \bibnamefont {Shi}}, \bibinfo {author} {\bibfnamefont
  {C.}~\bibnamefont {Burch}}, \bibinfo {author} {\bibfnamefont {L.~E.}\
  \bibnamefont {Anderson}}, \bibinfo {author} {\bibfnamefont {A.~T.}\
  \bibnamefont {Pierce}}, \bibinfo {author} {\bibfnamefont {Y.}~\bibnamefont
  {Xie}}, \bibinfo {author} {\bibfnamefont {A.}~\bibnamefont {Hamo}}, \bibinfo
  {author} {\bibfnamefont {U.}~\bibnamefont {Vool}}, \bibinfo {author}
  {\bibfnamefont {H.}~\bibnamefont {Zhang}}, \bibinfo {author} {\bibfnamefont
  {F.}~\bibnamefont {Casola}}, \bibinfo {author} {\bibfnamefont
  {T.}~\bibnamefont {Taniguchi}}, \bibinfo {author} {\bibfnamefont
  {K.}~\bibnamefont {Watanabe}}, \bibinfo {author} {\bibfnamefont {M.~M.}\
  \bibnamefont {Fogler}}, \bibinfo {author} {\bibfnamefont {P.}~\bibnamefont
  {Kim}}, \bibinfo {author} {\bibfnamefont {A.}~\bibnamefont {Yacoby}},\ and\
  \bibinfo {author} {\bibfnamefont {R.~L.}\ \bibnamefont {Walsworth}},\
  }\bibfield  {title} {\bibinfo {title} {Imaging viscous flow of the dirac
  fluid in graphene},\ }\href {https://doi.org/10.1038/s41586-020-2507-2}
  {\bibfield  {journal} {\bibinfo  {journal} {Nature}\ }\textbf {\bibinfo
  {volume} {583}},\ \bibinfo {pages} {537} (\bibinfo {year}
  {2020})}\BibitemShut {NoStop}%
\bibitem [{\citenamefont {Rondin}\ \emph {et~al.}(2013)\citenamefont {Rondin},
  \citenamefont {Tetienne}, \citenamefont {Rohart}, \citenamefont {Thiaville},
  \citenamefont {Hingant}, \citenamefont {Spinicelli}, \citenamefont {Roch},\
  and\ \citenamefont {Jacques}}]{rondin13}%
  \BibitemOpen
  \bibfield  {author} {\bibinfo {author} {\bibfnamefont {L.}~\bibnamefont
  {Rondin}}, \bibinfo {author} {\bibfnamefont {J.~P.}\ \bibnamefont
  {Tetienne}}, \bibinfo {author} {\bibfnamefont {S.}~\bibnamefont {Rohart}},
  \bibinfo {author} {\bibfnamefont {A.}~\bibnamefont {Thiaville}}, \bibinfo
  {author} {\bibfnamefont {T.}~\bibnamefont {Hingant}}, \bibinfo {author}
  {\bibfnamefont {P.}~\bibnamefont {Spinicelli}}, \bibinfo {author}
  {\bibfnamefont {J.~F.}\ \bibnamefont {Roch}},\ and\ \bibinfo {author}
  {\bibfnamefont {V.}~\bibnamefont {Jacques}},\ }\bibfield  {title} {\bibinfo
  {title} {Stray-field imaging of magnetic vortices with a single diamond
  spin},\ }\href {https://doi.org/10.1038/ncomms3279} {\bibfield  {journal}
  {\bibinfo  {journal} {Nat. Commun.}\ }\textbf {\bibinfo {volume} {4}},\
  \bibinfo {pages} {2279} (\bibinfo {year} {2013})}\BibitemShut {NoStop}%
\bibitem [{\citenamefont {Tetienne}\ \emph {et~al.}(2014)\citenamefont
  {Tetienne}, \citenamefont {Hingant}, \citenamefont {Kim}, \citenamefont
  {Diez}, \citenamefont {Adam}, \citenamefont {Garcia}, \citenamefont {Roch},
  \citenamefont {Rohart}, \citenamefont {Thiaville}, \citenamefont
  {Ravelosona},\ and\ \citenamefont {Jacques}}]{tetienne14}%
  \BibitemOpen
  \bibfield  {author} {\bibinfo {author} {\bibfnamefont {J.~P.}\ \bibnamefont
  {Tetienne}}, \bibinfo {author} {\bibfnamefont {T.}~\bibnamefont {Hingant}},
  \bibinfo {author} {\bibfnamefont {J.}~\bibnamefont {Kim}}, \bibinfo {author}
  {\bibfnamefont {L.~H.}\ \bibnamefont {Diez}}, \bibinfo {author}
  {\bibfnamefont {J.~P.}\ \bibnamefont {Adam}}, \bibinfo {author}
  {\bibfnamefont {K.}~\bibnamefont {Garcia}}, \bibinfo {author} {\bibfnamefont
  {J.~F.}\ \bibnamefont {Roch}}, \bibinfo {author} {\bibfnamefont
  {S.}~\bibnamefont {Rohart}}, \bibinfo {author} {\bibfnamefont
  {A.}~\bibnamefont {Thiaville}}, \bibinfo {author} {\bibfnamefont
  {D.}~\bibnamefont {Ravelosona}},\ and\ \bibinfo {author} {\bibfnamefont
  {V.}~\bibnamefont {Jacques}},\ }\bibfield  {title} {\bibinfo {title}
  {Nanoscale imaging and control of domain-wall hopping with a nitrogen-vacancy
  center microscope},\ }\href {https://doi.org/10.1126/science.1250113}
  {\bibfield  {journal} {\bibinfo  {journal} {Science}\ }\textbf {\bibinfo
  {volume} {344}},\ \bibinfo {pages} {1366} (\bibinfo {year}
  {2014})}\BibitemShut {NoStop}%
\bibitem [{\citenamefont {Tetienne}\ \emph {et~al.}(2015)\citenamefont
  {Tetienne}, \citenamefont {Hingant}, \citenamefont {Martinez}, \citenamefont
  {Rohart}, \citenamefont {Thiaville}, \citenamefont {Diez}, \citenamefont
  {Garcia}, \citenamefont {Adam}, \citenamefont {Kim}, \citenamefont {Roch},
  \citenamefont {Miron}, \citenamefont {Gaudin}, \citenamefont {Vila},
  \citenamefont {Ocker}, \citenamefont {Ravelosona},\ and\ \citenamefont
  {Jacques}}]{tetienne15}%
  \BibitemOpen
  \bibfield  {author} {\bibinfo {author} {\bibfnamefont {J.~P.}\ \bibnamefont
  {Tetienne}}, \bibinfo {author} {\bibfnamefont {T.}~\bibnamefont {Hingant}},
  \bibinfo {author} {\bibfnamefont {L.~J.}\ \bibnamefont {Martinez}}, \bibinfo
  {author} {\bibfnamefont {S.}~\bibnamefont {Rohart}}, \bibinfo {author}
  {\bibfnamefont {A.}~\bibnamefont {Thiaville}}, \bibinfo {author}
  {\bibfnamefont {L.~H.}\ \bibnamefont {Diez}}, \bibinfo {author}
  {\bibfnamefont {K.}~\bibnamefont {Garcia}}, \bibinfo {author} {\bibfnamefont
  {J.~P.}\ \bibnamefont {Adam}}, \bibinfo {author} {\bibfnamefont {J.~V.}\
  \bibnamefont {Kim}}, \bibinfo {author} {\bibfnamefont {J.~F.}\ \bibnamefont
  {Roch}}, \bibinfo {author} {\bibfnamefont {I.~M.}\ \bibnamefont {Miron}},
  \bibinfo {author} {\bibfnamefont {G.}~\bibnamefont {Gaudin}}, \bibinfo
  {author} {\bibfnamefont {L.}~\bibnamefont {Vila}}, \bibinfo {author}
  {\bibfnamefont {B.}~\bibnamefont {Ocker}}, \bibinfo {author} {\bibfnamefont
  {D.}~\bibnamefont {Ravelosona}},\ and\ \bibinfo {author} {\bibfnamefont
  {V.}~\bibnamefont {Jacques}},\ }\bibfield  {title} {\bibinfo {title} {The
  nature of domain walls in ultrathin ferromagnets revealed by scanning
  nanomagnetometry},\ }\href {https://doi.org/10.1038/ncomms7733} {\bibfield
  {journal} {\bibinfo  {journal} {Nat. Commun.}\ }\textbf {\bibinfo {volume}
  {6}},\ \bibinfo {pages} {6733} (\bibinfo {year} {2015})}\BibitemShut
  {NoStop}%
\bibitem [{\citenamefont {Dussaux}\ \emph {et~al.}(2016)\citenamefont
  {Dussaux}, \citenamefont {Schoenherr}, \citenamefont {Koumpouras},
  \citenamefont {Chico}, \citenamefont {Chang}, \citenamefont {Lorenzelli},
  \citenamefont {Kanazawa}, \citenamefont {Tokura}, \citenamefont {Garst},
  \citenamefont {Bergman}, \citenamefont {Degen},\ and\ \citenamefont
  {Meier}}]{dussaux16}%
  \BibitemOpen
  \bibfield  {author} {\bibinfo {author} {\bibfnamefont {A.}~\bibnamefont
  {Dussaux}}, \bibinfo {author} {\bibfnamefont {P.}~\bibnamefont {Schoenherr}},
  \bibinfo {author} {\bibfnamefont {K.}~\bibnamefont {Koumpouras}}, \bibinfo
  {author} {\bibfnamefont {J.}~\bibnamefont {Chico}}, \bibinfo {author}
  {\bibfnamefont {K.}~\bibnamefont {Chang}}, \bibinfo {author} {\bibfnamefont
  {L.}~\bibnamefont {Lorenzelli}}, \bibinfo {author} {\bibfnamefont
  {N.}~\bibnamefont {Kanazawa}}, \bibinfo {author} {\bibfnamefont
  {Y.}~\bibnamefont {Tokura}}, \bibinfo {author} {\bibfnamefont
  {M.}~\bibnamefont {Garst}}, \bibinfo {author} {\bibfnamefont
  {A.}~\bibnamefont {Bergman}}, \bibinfo {author} {\bibfnamefont {C.~L.}\
  \bibnamefont {Degen}},\ and\ \bibinfo {author} {\bibfnamefont
  {D.}~\bibnamefont {Meier}},\ }\bibfield  {title} {\bibinfo {title} {Local
  dynamics of topological magnetic defects in the itinerant helimagnet fege},\
  }\href {https://doi.org/10.1038/ncomms12430} {\bibfield  {journal} {\bibinfo
  {journal} {Nature Communications}\ }\textbf {\bibinfo {volume} {7}},\
  \bibinfo {pages} {12430} (\bibinfo {year} {2016})}\BibitemShut {NoStop}%
\bibitem [{\citenamefont {Gross}\ \emph {et~al.}(2018)\citenamefont {Gross},
  \citenamefont {Akhtar}, \citenamefont {Hrabec}, \citenamefont {Sampaio},
  \citenamefont {Martinez}, \citenamefont {Chouaieb}, \citenamefont {Shields},
  \citenamefont {Maletinsky}, \citenamefont {Thiaville}, \citenamefont
  {Rohart},\ and\ \citenamefont {Jacques}}]{gross18}%
  \BibitemOpen
  \bibfield  {author} {\bibinfo {author} {\bibfnamefont {I.}~\bibnamefont
  {Gross}}, \bibinfo {author} {\bibfnamefont {W.}~\bibnamefont {Akhtar}},
  \bibinfo {author} {\bibfnamefont {A.}~\bibnamefont {Hrabec}}, \bibinfo
  {author} {\bibfnamefont {J.}~\bibnamefont {Sampaio}}, \bibinfo {author}
  {\bibfnamefont {L.~J.}\ \bibnamefont {Martinez}}, \bibinfo {author}
  {\bibfnamefont {S.}~\bibnamefont {Chouaieb}}, \bibinfo {author}
  {\bibfnamefont {B.~J.}\ \bibnamefont {Shields}}, \bibinfo {author}
  {\bibfnamefont {P.}~\bibnamefont {Maletinsky}}, \bibinfo {author}
  {\bibfnamefont {A.}~\bibnamefont {Thiaville}}, \bibinfo {author}
  {\bibfnamefont {S.}~\bibnamefont {Rohart}},\ and\ \bibinfo {author}
  {\bibfnamefont {V.}~\bibnamefont {Jacques}},\ }\bibfield  {title} {\bibinfo
  {title} {Skyrmion morphology in ultrathin magnetic films},\ }\href
  {https://doi.org/10.1103/PhysRevMaterials.2.024406} {\bibfield  {journal}
  {\bibinfo  {journal} {Phys. Rev. Materials}\ }\textbf {\bibinfo {volume}
  {2}},\ \bibinfo {pages} {024406} (\bibinfo {year} {2018})}\BibitemShut
  {NoStop}%
\bibitem [{\citenamefont {Jenkins}\ \emph {et~al.}(2019)\citenamefont
  {Jenkins}, \citenamefont {Pelliccione}, \citenamefont {Yu}, \citenamefont
  {Ma}, \citenamefont {Li}, \citenamefont {Wang},\ and\ \citenamefont
  {Jayich}}]{jenkins19}%
  \BibitemOpen
  \bibfield  {author} {\bibinfo {author} {\bibfnamefont {A.}~\bibnamefont
  {Jenkins}}, \bibinfo {author} {\bibfnamefont {M.}~\bibnamefont
  {Pelliccione}}, \bibinfo {author} {\bibfnamefont {G.}~\bibnamefont {Yu}},
  \bibinfo {author} {\bibfnamefont {X.}~\bibnamefont {Ma}}, \bibinfo {author}
  {\bibfnamefont {X.}~\bibnamefont {Li}}, \bibinfo {author} {\bibfnamefont
  {K.~L.}\ \bibnamefont {Wang}},\ and\ \bibinfo {author} {\bibfnamefont
  {A.~C.~B.}\ \bibnamefont {Jayich}},\ }\bibfield  {title} {\bibinfo {title}
  {Single-spin sensing of domain-wall structure and dynamics in a thin-film
  skyrmion host},\ }\href {https://doi.org/10.1103/PhysRevMaterials.3.083801}
  {\bibfield  {journal} {\bibinfo  {journal} {Phys. Rev. Materials}\ }\textbf
  {\bibinfo {volume} {3}},\ \bibinfo {pages} {083801} (\bibinfo {year}
  {2019})}\BibitemShut {NoStop}%
\bibitem [{\citenamefont {Thiel}\ \emph {et~al.}(2019)\citenamefont {Thiel},
  \citenamefont {Wang}, \citenamefont {Tschudin}, \citenamefont {Rohner},
  \citenamefont {Gutierrez-lezama}, \citenamefont {Ubrig}, \citenamefont
  {Gibertini}, \citenamefont {Giannini}, \citenamefont {Morpurgo},\ and\
  \citenamefont {Maletinsky}}]{thiel19}%
  \BibitemOpen
  \bibfield  {author} {\bibinfo {author} {\bibfnamefont {L.}~\bibnamefont
  {Thiel}}, \bibinfo {author} {\bibfnamefont {Z.}~\bibnamefont {Wang}},
  \bibinfo {author} {\bibfnamefont {M.~A.}\ \bibnamefont {Tschudin}}, \bibinfo
  {author} {\bibfnamefont {D.}~\bibnamefont {Rohner}}, \bibinfo {author}
  {\bibfnamefont {I.}~\bibnamefont {Gutierrez-lezama}}, \bibinfo {author}
  {\bibfnamefont {N.}~\bibnamefont {Ubrig}}, \bibinfo {author} {\bibfnamefont
  {M.}~\bibnamefont {Gibertini}}, \bibinfo {author} {\bibfnamefont
  {E.}~\bibnamefont {Giannini}}, \bibinfo {author} {\bibfnamefont {A.~F.}\
  \bibnamefont {Morpurgo}},\ and\ \bibinfo {author} {\bibfnamefont
  {P.}~\bibnamefont {Maletinsky}},\ }\bibfield  {title} {\bibinfo {title}
  {Probing magnetism in {{2D}} materials at the nanoscale with single-spin
  microscopy},\ }\href {https://doi.org/10.1126/science.aav6926} {\bibfield
  {journal} {\bibinfo  {journal} {Science}\ }\textbf {\bibinfo {volume}
  {364}},\ \bibinfo {pages} {973} (\bibinfo {year} {2019})}\BibitemShut
  {NoStop}%
\bibitem [{\citenamefont {Sun}\ \emph {et~al.}(2021)\citenamefont {Sun},
  \citenamefont {Song}, \citenamefont {Anderson}, \citenamefont {Brunner},
  \citenamefont {Forster}, \citenamefont {Shalomayeva}, \citenamefont
  {Taniguchi}, \citenamefont {Watanabe}, \citenamefont {Grafe}, \citenamefont
  {Stohr}, \citenamefont {Xu},\ and\ \citenamefont {Wrachtrup}}]{sun21}%
  \BibitemOpen
  \bibfield  {author} {\bibinfo {author} {\bibfnamefont {Q.}~\bibnamefont
  {Sun}}, \bibinfo {author} {\bibfnamefont {T.}~\bibnamefont {Song}}, \bibinfo
  {author} {\bibfnamefont {E.}~\bibnamefont {Anderson}}, \bibinfo {author}
  {\bibfnamefont {A.}~\bibnamefont {Brunner}}, \bibinfo {author} {\bibfnamefont
  {J.}~\bibnamefont {Forster}}, \bibinfo {author} {\bibfnamefont
  {T.}~\bibnamefont {Shalomayeva}}, \bibinfo {author} {\bibfnamefont
  {T.}~\bibnamefont {Taniguchi}}, \bibinfo {author} {\bibfnamefont
  {K.}~\bibnamefont {Watanabe}}, \bibinfo {author} {\bibfnamefont
  {J.}~\bibnamefont {Grafe}}, \bibinfo {author} {\bibfnamefont
  {R.}~\bibnamefont {Stohr}}, \bibinfo {author} {\bibfnamefont
  {X.}~\bibnamefont {Xu}},\ and\ \bibinfo {author} {\bibfnamefont
  {J.}~\bibnamefont {Wrachtrup}},\ }\bibfield  {title} {\bibinfo {title}
  {Magnetic domains and domain wall pinning in atomically thin crbr$_3$
  revealed by nanoscale imaging},\ }\href
  {https://doi.org/10.1038/s41467-021-22239-4} {\bibfield  {journal} {\bibinfo
  {journal} {Nature Communications}\ }\textbf {\bibinfo {volume} {12}},\
  \bibinfo {pages} {1989} (\bibinfo {year} {2021})}\BibitemShut {NoStop}%
\bibitem [{\citenamefont {Fabre}\ \emph {et~al.}(2021)\citenamefont {Fabre},
  \citenamefont {Finco}, \citenamefont {Purbawati}, \citenamefont {Hadj-Azzem},
  \citenamefont {Rougemaille}, \citenamefont {Coraux}, \citenamefont {Philip},\
  and\ \citenamefont {Jacques}}]{fabre21}%
  \BibitemOpen
  \bibfield  {author} {\bibinfo {author} {\bibfnamefont {F.}~\bibnamefont
  {Fabre}}, \bibinfo {author} {\bibfnamefont {A.}~\bibnamefont {Finco}},
  \bibinfo {author} {\bibfnamefont {A.}~\bibnamefont {Purbawati}}, \bibinfo
  {author} {\bibfnamefont {A.}~\bibnamefont {Hadj-Azzem}}, \bibinfo {author}
  {\bibfnamefont {N.}~\bibnamefont {Rougemaille}}, \bibinfo {author}
  {\bibfnamefont {J.}~\bibnamefont {Coraux}}, \bibinfo {author} {\bibfnamefont
  {I.}~\bibnamefont {Philip}},\ and\ \bibinfo {author} {\bibfnamefont
  {V.}~\bibnamefont {Jacques}},\ }\bibfield  {title} {\bibinfo {title}
  {Characterization of room-temperature in-plane magnetization in thin flakes
  of crte$_2$ with a single-spin magnetometer},\ }\href
  {https://doi.org/10.1103/PhysRevMaterials.5.034008} {\bibfield  {journal}
  {\bibinfo  {journal} {Phys. Rev. Materials}\ }\textbf {\bibinfo {volume}
  {5}},\ \bibinfo {pages} {034008} (\bibinfo {year} {2021})}\BibitemShut
  {NoStop}%
\bibitem [{\citenamefont {V\'elez}\ \emph {et~al.}(2019)\citenamefont
  {V\'elez}, \citenamefont {Schaab}, \citenamefont {W\"ornle}, \citenamefont
  {M\"uller}, \citenamefont {Gradauskaite}, \citenamefont {Welter},
  \citenamefont {Gutgsell}, \citenamefont {Nistor}, \citenamefont {Degen},
  \citenamefont {Trassin}, \citenamefont {Fiebig},\ and\ \citenamefont
  {Gambardella}}]{velez19}%
  \BibitemOpen
  \bibfield  {author} {\bibinfo {author} {\bibfnamefont {S.}~\bibnamefont
  {V\'elez}}, \bibinfo {author} {\bibfnamefont {J.}~\bibnamefont {Schaab}},
  \bibinfo {author} {\bibfnamefont {M.~S.}\ \bibnamefont {W\"ornle}}, \bibinfo
  {author} {\bibfnamefont {M.}~\bibnamefont {M\"uller}}, \bibinfo {author}
  {\bibfnamefont {E.}~\bibnamefont {Gradauskaite}}, \bibinfo {author}
  {\bibfnamefont {P.}~\bibnamefont {Welter}}, \bibinfo {author} {\bibfnamefont
  {C.}~\bibnamefont {Gutgsell}}, \bibinfo {author} {\bibfnamefont
  {C.}~\bibnamefont {Nistor}}, \bibinfo {author} {\bibfnamefont {C.~L.}\
  \bibnamefont {Degen}}, \bibinfo {author} {\bibfnamefont {M.}~\bibnamefont
  {Trassin}}, \bibinfo {author} {\bibfnamefont {M.}~\bibnamefont {Fiebig}},\
  and\ \bibinfo {author} {\bibfnamefont {P.}~\bibnamefont {Gambardella}},\
  }\bibfield  {title} {\bibinfo {title} {High-speed domain wall racetracks in a
  magnetic insulator},\ }\href {https://doi.org/10.1038/s41467-019-12676-7}
  {\bibfield  {journal} {\bibinfo  {journal} {Nature Communications}\ }\textbf
  {\bibinfo {volume} {10}},\ \bibinfo {pages} {4750} (\bibinfo {year}
  {2019})}\BibitemShut {NoStop}%
\bibitem [{\citenamefont {Wornle}\ \emph {et~al.}(2021)\citenamefont {Wornle},
  \citenamefont {Welter}, \citenamefont {Giraldo}, \citenamefont {Lottermoser},
  \citenamefont {Fiebig}, \citenamefont {Gambardella},\ and\ \citenamefont
  {Degen}}]{wornle21}%
  \BibitemOpen
  \bibfield  {author} {\bibinfo {author} {\bibfnamefont {M.~S.}\ \bibnamefont
  {Wornle}}, \bibinfo {author} {\bibfnamefont {P.}~\bibnamefont {Welter}},
  \bibinfo {author} {\bibfnamefont {M.}~\bibnamefont {Giraldo}}, \bibinfo
  {author} {\bibfnamefont {T.}~\bibnamefont {Lottermoser}}, \bibinfo {author}
  {\bibfnamefont {M.}~\bibnamefont {Fiebig}}, \bibinfo {author} {\bibfnamefont
  {P.}~\bibnamefont {Gambardella}},\ and\ \bibinfo {author} {\bibfnamefont
  {C.~L.}\ \bibnamefont {Degen}},\ }\bibfield  {title} {\bibinfo {title}
  {Coexistence of bloch and neel walls in a collinear antiferromagnet},\ }\href
  {https://doi.org/10.1103/PhysRevB.103.094426} {\bibfield  {journal} {\bibinfo
   {journal} {Phys. Rev. B}\ }\textbf {\bibinfo {volume} {103}},\ \bibinfo
  {pages} {094426} (\bibinfo {year} {2021})}\BibitemShut {NoStop}%
\bibitem [{\citenamefont {Hedrich}\ \emph {et~al.}(2021)\citenamefont
  {Hedrich}, \citenamefont {Wagner}, \citenamefont {Pylypovskyi}, \citenamefont
  {Shields}, \citenamefont {Kosub}, \citenamefont {Sheka}, \citenamefont
  {Makarov},\ and\ \citenamefont {Maletinsky}}]{hedrich21}%
  \BibitemOpen
  \bibfield  {author} {\bibinfo {author} {\bibfnamefont {N.}~\bibnamefont
  {Hedrich}}, \bibinfo {author} {\bibfnamefont {K.}~\bibnamefont {Wagner}},
  \bibinfo {author} {\bibfnamefont {O.~V.}\ \bibnamefont {Pylypovskyi}},
  \bibinfo {author} {\bibfnamefont {B.~J.}\ \bibnamefont {Shields}}, \bibinfo
  {author} {\bibfnamefont {T.}~\bibnamefont {Kosub}}, \bibinfo {author}
  {\bibfnamefont {D.~D.}\ \bibnamefont {Sheka}}, \bibinfo {author}
  {\bibfnamefont {D.}~\bibnamefont {Makarov}},\ and\ \bibinfo {author}
  {\bibfnamefont {P.}~\bibnamefont {Maletinsky}},\ }\bibfield  {title}
  {\bibinfo {title} {Nanoscale mechanics of antiferromagnetic domain walls},\
  }\href {https://doi.org/10.1038/s41567-020-01157-0} {\bibfield  {journal}
  {\bibinfo  {journal} {Nature Physics}\ }\textbf {\bibinfo {volume} {17}},\
  \bibinfo {pages} {064007} (\bibinfo {year} {2021})}\BibitemShut {NoStop}%
\bibitem [{\citenamefont {Wolfe}\ \emph {et~al.}(2014)\citenamefont {Wolfe},
  \citenamefont {Bhallamudi}, \citenamefont {Wang}, \citenamefont {Du},
  \citenamefont {Manuilov}, \citenamefont {Teeling-Smith}, \citenamefont
  {Berger}, \citenamefont {Adur}, \citenamefont {Yang},\ and\ \citenamefont
  {Hammel}}]{wolfe14}%
  \BibitemOpen
  \bibfield  {author} {\bibinfo {author} {\bibfnamefont {C.~S.}\ \bibnamefont
  {Wolfe}}, \bibinfo {author} {\bibfnamefont {V.~P.}\ \bibnamefont
  {Bhallamudi}}, \bibinfo {author} {\bibfnamefont {H.~L.}\ \bibnamefont
  {Wang}}, \bibinfo {author} {\bibfnamefont {C.~H.}\ \bibnamefont {Du}},
  \bibinfo {author} {\bibfnamefont {S.}~\bibnamefont {Manuilov}}, \bibinfo
  {author} {\bibfnamefont {R.~M.}\ \bibnamefont {Teeling-Smith}}, \bibinfo
  {author} {\bibfnamefont {A.~J.}\ \bibnamefont {Berger}}, \bibinfo {author}
  {\bibfnamefont {R.}~\bibnamefont {Adur}}, \bibinfo {author} {\bibfnamefont
  {F.~Y.}\ \bibnamefont {Yang}},\ and\ \bibinfo {author} {\bibfnamefont
  {P.~C.}\ \bibnamefont {Hammel}},\ }\bibfield  {title} {\bibinfo {title}
  {Off-resonant manipulation of spins in diamond via precessing magnetization
  of a proximal ferromagnet},\ }\href
  {https://doi.org/10.1103/PhysRevB.89.180406} {\bibfield  {journal} {\bibinfo
  {journal} {Phys. Rev. B}\ }\textbf {\bibinfo {volume} {89}},\ \bibinfo
  {pages} {180406} (\bibinfo {year} {2014})}\BibitemShut {NoStop}%
\bibitem [{\citenamefont {der Sar}\ \emph {et~al.}(2015)\citenamefont {der
  Sar}, \citenamefont {Casola}, \citenamefont {Walsworth},\ and\ \citenamefont
  {Yacoby}}]{vandersar15}%
  \BibitemOpen
  \bibfield  {author} {\bibinfo {author} {\bibfnamefont {T.~V.}\ \bibnamefont
  {der Sar}}, \bibinfo {author} {\bibfnamefont {F.}~\bibnamefont {Casola}},
  \bibinfo {author} {\bibfnamefont {R.}~\bibnamefont {Walsworth}},\ and\
  \bibinfo {author} {\bibfnamefont {A.}~\bibnamefont {Yacoby}},\ }\bibfield
  {title} {\bibinfo {title} {Nanometre-scale probing of spin waves using single
  electron spins},\ }\href {https://doi.org/10.1038/ncomms8886} {\bibfield
  {journal} {\bibinfo  {journal} {Nat. Commun.}\ }\textbf {\bibinfo {volume}
  {6}},\ \bibinfo {pages} {7886} (\bibinfo {year} {2015})}\BibitemShut
  {NoStop}%
\bibitem [{\citenamefont {Bertelli}\ \emph {et~al.}(2020)\citenamefont
  {Bertelli}, \citenamefont {Carmiggelt}, \citenamefont {Yu}, \citenamefont
  {Simon}, \citenamefont {Pothoven}, \citenamefont {Bauer}, \citenamefont
  {Blanter}, \citenamefont {Aarts},\ and\ \citenamefont {der
  sar}}]{bertelli20}%
  \BibitemOpen
  \bibfield  {author} {\bibinfo {author} {\bibfnamefont {I.}~\bibnamefont
  {Bertelli}}, \bibinfo {author} {\bibfnamefont {J.~J.}\ \bibnamefont
  {Carmiggelt}}, \bibinfo {author} {\bibfnamefont {T.}~\bibnamefont {Yu}},
  \bibinfo {author} {\bibfnamefont {B.~G.}\ \bibnamefont {Simon}}, \bibinfo
  {author} {\bibfnamefont {C.~C.}\ \bibnamefont {Pothoven}}, \bibinfo {author}
  {\bibfnamefont {G.~E.~W.}\ \bibnamefont {Bauer}}, \bibinfo {author}
  {\bibfnamefont {Y.~M.}\ \bibnamefont {Blanter}}, \bibinfo {author}
  {\bibfnamefont {J.}~\bibnamefont {Aarts}},\ and\ \bibinfo {author}
  {\bibfnamefont {T.~V.}\ \bibnamefont {der sar}},\ }\bibfield  {title}
  {\bibinfo {title} {Magnetic resonance imaging of spin-wave transport and
  interference in a magnetic insulator},\ }\href
  {https://doi.org/10.1126/sciadv.abd3556} {\bibfield  {journal} {\bibinfo
  {journal} {Sci. Adv.}\ }\textbf {\bibinfo {volume} {6}},\ \bibinfo {pages}
  {eabd3556} (\bibinfo {year} {2020})}\BibitemShut {NoStop}%
\bibitem [{\citenamefont {Epstein}\ \emph {et~al.}(2005)\citenamefont
  {Epstein}, \citenamefont {Mendoza}, \citenamefont {Kato},\ and\ \citenamefont
  {Awschalom}}]{epstein05}%
  \BibitemOpen
  \bibfield  {author} {\bibinfo {author} {\bibfnamefont {R.~J.}\ \bibnamefont
  {Epstein}}, \bibinfo {author} {\bibfnamefont {F.~M.}\ \bibnamefont
  {Mendoza}}, \bibinfo {author} {\bibfnamefont {Y.~K.}\ \bibnamefont {Kato}},\
  and\ \bibinfo {author} {\bibfnamefont {D.~D.}\ \bibnamefont {Awschalom}},\
  }\bibfield  {title} {\bibinfo {title} {Anisotropic interactions of a single
  spin and dark-spin spectroscopy in diamond},\ }\href
  {https://doi.org/10.1038/nphys141} {\bibfield  {journal} {\bibinfo  {journal}
  {Nat. Phys.}\ }\textbf {\bibinfo {volume} {1}},\ \bibinfo {eid} {94}
  (\bibinfo {year} {2005})}\BibitemShut {NoStop}%
\bibitem [{\citenamefont {Budker}\ and\ \citenamefont
  {Romalis}(2007)}]{budker07}%
  \BibitemOpen
  \bibfield  {author} {\bibinfo {author} {\bibfnamefont {D.}~\bibnamefont
  {Budker}}\ and\ \bibinfo {author} {\bibfnamefont {M.}~\bibnamefont
  {Romalis}},\ }\bibfield  {title} {\bibinfo {title} {Optical magnetometry},\
  }\href {https://doi.org/10.1038/nphys566} {\bibfield  {journal} {\bibinfo
  {journal} {Nat. Phys.}\ }\textbf {\bibinfo {volume} {3}},\ \bibinfo {eid}
  {227} (\bibinfo {year} {2007})}\BibitemShut {NoStop}%
\bibitem [{\citenamefont {Schirhagl}\ \emph {et~al.}(2014)\citenamefont
  {Schirhagl}, \citenamefont {Chang}, \citenamefont {Loretz},\ and\
  \citenamefont {Degen}}]{schirhagl14}%
  \BibitemOpen
  \bibfield  {author} {\bibinfo {author} {\bibfnamefont {R.}~\bibnamefont
  {Schirhagl}}, \bibinfo {author} {\bibfnamefont {K.}~\bibnamefont {Chang}},
  \bibinfo {author} {\bibfnamefont {M.}~\bibnamefont {Loretz}},\ and\ \bibinfo
  {author} {\bibfnamefont {C.~L.}\ \bibnamefont {Degen}},\ }\bibfield  {title}
  {\bibinfo {title} {Nitrogen-vacancy centers in diamond: Nanoscale sensors for
  physics and biology},\ }\href
  {https://doi.org/10.1146/annurev-physchem-040513-103659} {\bibfield
  {journal} {\bibinfo  {journal} {Annu. Rev. Phys. Chem.}\ }\textbf {\bibinfo
  {volume} {65}},\ \bibinfo {pages} {83} (\bibinfo {year} {2014})}\BibitemShut
  {NoStop}%
\bibitem [{\citenamefont {Dreau}\ \emph {et~al.}(2011)\citenamefont {Dreau},
  \citenamefont {Lesik}, \citenamefont {Rondin}, \citenamefont {Spinicelli},
  \citenamefont {Arcizet}, \citenamefont {Roch},\ and\ \citenamefont
  {Jacques}}]{dreau11}%
  \BibitemOpen
  \bibfield  {author} {\bibinfo {author} {\bibfnamefont {A.}~\bibnamefont
  {Dreau}}, \bibinfo {author} {\bibfnamefont {M.}~\bibnamefont {Lesik}},
  \bibinfo {author} {\bibfnamefont {L.}~\bibnamefont {Rondin}}, \bibinfo
  {author} {\bibfnamefont {P.}~\bibnamefont {Spinicelli}}, \bibinfo {author}
  {\bibfnamefont {O.}~\bibnamefont {Arcizet}}, \bibinfo {author} {\bibfnamefont
  {J.~F.}\ \bibnamefont {Roch}},\ and\ \bibinfo {author} {\bibfnamefont
  {V.}~\bibnamefont {Jacques}},\ }\bibfield  {title} {\bibinfo {title}
  {Avoiding power broadening in optically detected magnetic resonance of single
  nv defects for enhanced dc magnetic field sensitivity},\ }\href
  {https://doi.org/10.1103/PhysRevB.84.195204} {\bibfield  {journal} {\bibinfo
  {journal} {Phys. Rev. B}\ }\textbf {\bibinfo {volume} {84}},\ \bibinfo
  {pages} {195204} (\bibinfo {year} {2011})}\BibitemShut {NoStop}%
\bibitem [{\citenamefont {Neu}\ \emph {et~al.}(2014)\citenamefont {Neu},
  \citenamefont {Appel}, \citenamefont {Ganzhorn}, \citenamefont
  {Miguel-Sanchez}, \citenamefont {Lesik}, \citenamefont {Mille}, \citenamefont
  {Jacques}, \citenamefont {Tallaire}, \citenamefont {Achard},\ and\
  \citenamefont {Maletinsky}}]{neu14}%
  \BibitemOpen
  \bibfield  {author} {\bibinfo {author} {\bibfnamefont {E.}~\bibnamefont
  {Neu}}, \bibinfo {author} {\bibfnamefont {P.}~\bibnamefont {Appel}}, \bibinfo
  {author} {\bibfnamefont {M.}~\bibnamefont {Ganzhorn}}, \bibinfo {author}
  {\bibfnamefont {J.}~\bibnamefont {Miguel-Sanchez}}, \bibinfo {author}
  {\bibfnamefont {M.}~\bibnamefont {Lesik}}, \bibinfo {author} {\bibfnamefont
  {V.}~\bibnamefont {Mille}}, \bibinfo {author} {\bibfnamefont
  {V.}~\bibnamefont {Jacques}}, \bibinfo {author} {\bibfnamefont
  {A.}~\bibnamefont {Tallaire}}, \bibinfo {author} {\bibfnamefont
  {J.}~\bibnamefont {Achard}},\ and\ \bibinfo {author} {\bibfnamefont
  {P.}~\bibnamefont {Maletinsky}},\ }\bibfield  {title} {\bibinfo {title}
  {Photonic nano-structures on (111)-oriented diamond},\ }\href
  {https://doi.org/10.1063/1.4871580} {\bibfield  {journal} {\bibinfo
  {journal} {Appl. Phys. Lett.}\ }\textbf {\bibinfo {volume} {104}},\ \bibinfo
  {pages} {153108} (\bibinfo {year} {2014})}\BibitemShut {NoStop}%
\bibitem [{\citenamefont {Rohner}\ \emph {et~al.}(2019)\citenamefont {Rohner},
  \citenamefont {Happacher}, \citenamefont {Reiser}, \citenamefont {Tschudin},
  \citenamefont {Tallaire}, \citenamefont {Achard}, \citenamefont {Shields},\
  and\ \citenamefont {Maletinsky}}]{rohner19}%
  \BibitemOpen
  \bibfield  {author} {\bibinfo {author} {\bibfnamefont {D.}~\bibnamefont
  {Rohner}}, \bibinfo {author} {\bibfnamefont {J.}~\bibnamefont {Happacher}},
  \bibinfo {author} {\bibfnamefont {P.}~\bibnamefont {Reiser}}, \bibinfo
  {author} {\bibfnamefont {M.~A.}\ \bibnamefont {Tschudin}}, \bibinfo {author}
  {\bibfnamefont {A.}~\bibnamefont {Tallaire}}, \bibinfo {author}
  {\bibfnamefont {J.}~\bibnamefont {Achard}}, \bibinfo {author} {\bibfnamefont
  {B.~J.}\ \bibnamefont {Shields}},\ and\ \bibinfo {author} {\bibfnamefont
  {P.}~\bibnamefont {Maletinsky}},\ }\bibfield  {title} {\bibinfo {title}
  {(111)-oriented single crystal diamond tips for nanoscale scanning probe
  imaging of out-of-plane magnetic fields},\ }\href
  {https://doi.org/10.1063/1.5127101} {\bibfield  {journal} {\bibinfo
  {journal} {Appl. Phys. Lett.}\ }\textbf {\bibinfo {volume} {115}},\ \bibinfo
  {pages} {192401} (\bibinfo {year} {2019})}\BibitemShut {NoStop}%
\bibitem [{\citenamefont {{QZabre Ltd.}}()}]{qzabre}%
  \BibitemOpen
  \bibfield  {author} {\bibinfo {author} {\bibnamefont {{QZabre Ltd.}}},\
  }\bibfield  {title} {\bibinfo {title} {{https://qzabre.com}},\ }\href
  {https://qzabre.com} {\ }\BibitemShut {NoStop}%
\bibitem [{\citenamefont {Rondin}\ \emph {et~al.}(2010)\citenamefont {Rondin},
  \citenamefont {Dantelle}, \citenamefont {Slablab}, \citenamefont {Grosshans},
  \citenamefont {Treussart}, \citenamefont {Bergonzo}, \citenamefont
  {Perruchas}, \citenamefont {Gacoin}, \citenamefont {Chaigneau}, \citenamefont
  {Chang}, \citenamefont {Jacques},\ and\ \citenamefont {Roch}}]{rondin10}%
  \BibitemOpen
  \bibfield  {author} {\bibinfo {author} {\bibfnamefont {L.}~\bibnamefont
  {Rondin}}, \bibinfo {author} {\bibfnamefont {G.}~\bibnamefont {Dantelle}},
  \bibinfo {author} {\bibfnamefont {A.}~\bibnamefont {Slablab}}, \bibinfo
  {author} {\bibfnamefont {F.}~\bibnamefont {Grosshans}}, \bibinfo {author}
  {\bibfnamefont {F.}~\bibnamefont {Treussart}}, \bibinfo {author}
  {\bibfnamefont {P.}~\bibnamefont {Bergonzo}}, \bibinfo {author}
  {\bibfnamefont {S.}~\bibnamefont {Perruchas}}, \bibinfo {author}
  {\bibfnamefont {T.}~\bibnamefont {Gacoin}}, \bibinfo {author} {\bibfnamefont
  {M.}~\bibnamefont {Chaigneau}}, \bibinfo {author} {\bibfnamefont
  {H.}~\bibnamefont {Chang}}, \bibinfo {author} {\bibfnamefont
  {V.}~\bibnamefont {Jacques}},\ and\ \bibinfo {author} {\bibfnamefont
  {J.}~\bibnamefont {Roch}},\ }\bibfield  {title} {\bibinfo {title}
  {Surface-induced charge state conversion of nitrogen-vacancy defects in
  nanodiamonds},\ }\href {https://doi.org/10.1103/PhysRevB.82.115449}
  {\bibfield  {journal} {\bibinfo  {journal} {Phys. Rev. B}\ }\textbf {\bibinfo
  {volume} {82}},\ \bibinfo {pages} {115449} (\bibinfo {year}
  {2010})}\BibitemShut {NoStop}%
\bibitem [{\citenamefont {Wornle}(2021)}]{wornle21thesis}%
  \BibitemOpen
  \bibfield  {author} {\bibinfo {author} {\bibfnamefont {M.~S.}\ \bibnamefont
  {Wornle}},\ }\bibfield  {title} {\bibinfo {title} {Nanoscale scanning diamond
  magnetometry of antiferromagnets},\ }\href
  {https://doi.org/10.3929/ethz-b-000488794} {\bibfield  {journal} {\bibinfo
  {journal} {PhD Thesis, ETH Zurich}\ } (\bibinfo {year} {2021})}\BibitemShut
  {NoStop}%
\bibitem [{sup()}]{supplementary}%
  \BibitemOpen
  \href@noop {} {\bibinfo  {journal} {See Supplementary Materials accompanying
  this manuscript}\ }\BibitemShut {NoStop}%
\bibitem [{\citenamefont {Tetienne}\ \emph {et~al.}(2012)\citenamefont
  {Tetienne}, \citenamefont {Rondin}, \citenamefont {Spinicelli}, \citenamefont
  {Chipaux}, \citenamefont {Debuisschert}, \citenamefont {Roch},\ and\
  \citenamefont {Jacques}}]{tetienne12}%
  \BibitemOpen
\bibfield  {journal} {  }\bibfield  {author} {\bibinfo {author} {\bibfnamefont
  {J.}~\bibnamefont {Tetienne}}, \bibinfo {author} {\bibfnamefont
  {L.}~\bibnamefont {Rondin}}, \bibinfo {author} {\bibfnamefont
  {P.}~\bibnamefont {Spinicelli}}, \bibinfo {author} {\bibfnamefont
  {M.}~\bibnamefont {Chipaux}}, \bibinfo {author} {\bibfnamefont
  {T.}~\bibnamefont {Debuisschert}}, \bibinfo {author} {\bibfnamefont
  {J.}~\bibnamefont {Roch}},\ and\ \bibinfo {author} {\bibfnamefont
  {V.}~\bibnamefont {Jacques}},\ }\bibfield  {title} {\bibinfo {title}
  {Magnetic-field-dependent photodynamics of single nv defects in diamond: an
  application to qualitative all-optical magnetic imaging},\ }\href
  {http://dx.doi.org/10.1088/1367-2630/14/10/103033} {\bibfield  {journal}
  {\bibinfo  {journal} {New Journal of Physics}\ }\textbf {\bibinfo {volume}
  {14}},\ \bibinfo {pages} {103033} (\bibinfo {year} {2012})}\BibitemShut
  {NoStop}%
\bibitem [{\citenamefont {Mazalski}\ \emph {et~al.}(2020)\citenamefont
  {Mazalski}, \citenamefont {Anastaziak}, \citenamefont {Kuswik}, \citenamefont
  {Kurant}, \citenamefont {Sveklo},\ and\ \citenamefont
  {Maziewski}}]{mazalski20}%
  \BibitemOpen
  \bibfield  {author} {\bibinfo {author} {\bibfnamefont {P.}~\bibnamefont
  {Mazalski}}, \bibinfo {author} {\bibfnamefont {B.}~\bibnamefont
  {Anastaziak}}, \bibinfo {author} {\bibfnamefont {P.}~\bibnamefont {Kuswik}},
  \bibinfo {author} {\bibfnamefont {Z.}~\bibnamefont {Kurant}}, \bibinfo
  {author} {\bibfnamefont {I.}~\bibnamefont {Sveklo}},\ and\ \bibinfo {author}
  {\bibfnamefont {A.}~\bibnamefont {Maziewski}},\ }\bibfield  {title} {\bibinfo
  {title} {Demagnetization of an ultrathin co/nio bilayer with creation of
  submicrometer domains controlled by temperature-induced changes of magnetic
  anisotropy},\ }\href {https://doi.org/10.1016/j.jmmm.2020.166871} {\bibfield
  {journal} {\bibinfo  {journal} {Journal of Magnetism and Magnetic Materials}\
  }\textbf {\bibinfo {volume} {508}},\ \bibinfo {pages} {166871} (\bibinfo
  {year} {2020})}\BibitemShut {NoStop}%
\bibitem [{\citenamefont {Wu}\ \emph {et~al.}(2009)\citenamefont {Wu},
  \citenamefont {Choi}, \citenamefont {Scholl}, \citenamefont {Doran},
  \citenamefont {Arenholz}, \citenamefont {Wu}, \citenamefont {Won},
  \citenamefont {Hwang},\ and\ \citenamefont {Qiu}}]{wu09}%
  \BibitemOpen
  \bibfield  {author} {\bibinfo {author} {\bibfnamefont {J.}~\bibnamefont
  {Wu}}, \bibinfo {author} {\bibfnamefont {J.}~\bibnamefont {Choi}}, \bibinfo
  {author} {\bibfnamefont {A.}~\bibnamefont {Scholl}}, \bibinfo {author}
  {\bibfnamefont {A.}~\bibnamefont {Doran}}, \bibinfo {author} {\bibfnamefont
  {E.}~\bibnamefont {Arenholz}}, \bibinfo {author} {\bibfnamefont {Y.~Z.}\
  \bibnamefont {Wu}}, \bibinfo {author} {\bibfnamefont {C.}~\bibnamefont
  {Won}}, \bibinfo {author} {\bibfnamefont {C.}~\bibnamefont {Hwang}},\ and\
  \bibinfo {author} {\bibfnamefont {Z.~Q.}\ \bibnamefont {Qiu}},\ }\bibfield
  {title} {\bibinfo {title} {Element-specific study of the anomalous magnetic
  interlayer coupling across nio spacer layer in co/nio/fe/ag(001) using xmcd
  and xmld},\ }\href {https://doi.org/10.1103/PhysRevB.80.012409} {\bibfield
  {journal} {\bibinfo  {journal} {Phys. Rev. B}\ }\textbf {\bibinfo {volume}
  {80}},\ \bibinfo {pages} {012409} (\bibinfo {year} {2009})}\BibitemShut
  {NoStop}%
\bibitem [{\citenamefont {Miron}\ \emph {et~al.}(2011)\citenamefont {Miron},
  \citenamefont {Garello}, \citenamefont {Gaudin}, \citenamefont {Zermatten},
  \citenamefont {Costache}, \citenamefont {Auffret}, \citenamefont {Bandiera},
  \citenamefont {Rodmacq}, \citenamefont {Schuhl},\ and\ \citenamefont
  {Gambardella}}]{miron11}%
  \BibitemOpen
  \bibfield  {author} {\bibinfo {author} {\bibfnamefont {I.~M.}\ \bibnamefont
  {Miron}}, \bibinfo {author} {\bibfnamefont {K.}~\bibnamefont {Garello}},
  \bibinfo {author} {\bibfnamefont {G.}~\bibnamefont {Gaudin}}, \bibinfo
  {author} {\bibfnamefont {P.}~\bibnamefont {Zermatten}}, \bibinfo {author}
  {\bibfnamefont {M.~V.}\ \bibnamefont {Costache}}, \bibinfo {author}
  {\bibfnamefont {S.}~\bibnamefont {Auffret}}, \bibinfo {author} {\bibfnamefont
  {S.}~\bibnamefont {Bandiera}}, \bibinfo {author} {\bibfnamefont
  {B.}~\bibnamefont {Rodmacq}}, \bibinfo {author} {\bibfnamefont
  {A.}~\bibnamefont {Schuhl}},\ and\ \bibinfo {author} {\bibfnamefont
  {P.}~\bibnamefont {Gambardella}},\ }\bibfield  {title} {\bibinfo {title}
  {Perpendicular switching of a single ferromagnetic layer induced by in-plane
  current injection},\ }\href {https://doi.org/10.1038/nature10309} {\bibfield
  {journal} {\bibinfo  {journal} {Nature}\ }\textbf {\bibinfo {volume} {476}},\
  \bibinfo {pages} {189} (\bibinfo {year} {2011})}\BibitemShut {NoStop}%
\end{thebibliography}%
%\bibliography{C:/ETH/labview/library/library}
%\input{"manuscript.bbl"}

\end{document}